\documentclass[useAMS,usenatbib]{mnras}
\usepackage{caption}
\usepackage{graphicx}
\usepackage{lscape}
\usepackage[utf8x]{inputenc}
\usepackage{amsfonts}
\usepackage{tabularx}
\usepackage{longtable}
\usepackage{natbib}
\usepackage{graphicx}
\usepackage{epsfig}
\usepackage{verbatim}
\usepackage[flushleft]{threeparttable}
\include{journaldefs}
\usepackage{multirow, makecell}
\usepackage[caption=false]{subfig}
\usepackage{url}
\usepackage{textcomp}
\usepackage[squaren]{SIunits}
\usepackage{float}
\usepackage{aecompl}
\usepackage{adjustbox}
\usepackage{hyperref}
\usepackage{pdflscape}
\usepackage{hyperref}
\hypersetup{
    colorlinks=true,
    citecolor=blue,
    linkcolor=blue,
    filecolor=blue,      
    urlcolor=blue,
}

\def\arcs{$^{\prime\prime}\,$}
\def\mujybm   {${\rm \mu}$Jy\,beam$^{-1}$}
\def\mjybm   {${\rm m}$Jy\,beam$^{-1}$}

\begin{document}

   \title[HXAGN -- II. SEDs in the 5-45 GHz domain]{Hard-X-ray-selected Active Galactic Nuclei -- II. 
   Spectral Energy Distributions in the 5-45 GHz domain}

\author[Francesca Panessa]
{Francesca Panessa$^{1}$\thanks{Contact e-mail:\href{mailto:francesca.panessa@inaf.it}
{francesca.panessa@inaf.it}},
Elia Chiaraluce$^{2}$,
Gabriele Bruni$^{1}$,
Daniele Dallacasa$^{3,4}$,
Ari Laor$^{5}$,
\newauthor
Ranieri D. Baldi$^{4}$,
Ehud Behar$^{5}$,
Ian McHardy$^{6}$,
Francesco Tombesi$^{7,8,9,10,11}$,
Fausto Vagnetti$^{1,7}$
\\
\\
$^{1}$ INAF -- Istituto di Astrofisica e Planetologia Spaziali, via del Fosso del Cavaliere 100, I-00133 Roma, Italy\\
$^{2}$ AKKA Technologies Italia srl, via Tiburtina 1072/1074, 00156 Roma, Italy \\
$^{3}$ DiFA -- Dipartimento di Fisica e Astronomia, Università di Bologna, via P. Gobetti 93/2, I-40129 Bologna, Italy\\
$^{4}$ INAF -- Istituto di Radioastronomia, via Piero Gobetti 101, I-40129 Bologna, Italy\\
$^{5}$ Physics Department, The Technion, 32000, Haifa, Israel\\
$^{6}$ School of Physics and Astronomy, University of Southampton, Southampton, SO17 1BJ, UK\\
$^{7}$ Dipartimento di Fisica, Università di Roma “Tor Vergata”, Via della Ricerca Scientifica 1, I-00133 Roma, Italy \\
$^{8}$ INAF -- Astronomical Observatory of Rome, Via Frascati 33, I-00078 Monte Porzio Catone, Italy\\
$^{9}$ Department of Astronomy, University of Maryland, College Park MD 20742, USA\\
$^{10}$ X-ray Astrophysics Laboratory, NASA/Goddard Space Flight Center, Greenbelt, MD 20771, USA\\
$^{11}$ INFN -- Roma Tor Vergata, Via della Ricerca Scientifica 1, 00133, Roma, Italy
}

\pubyear{}

\newcommand{\Nh}{N_{\rm H}}
\newcommand{\Nii}{[N {\sc ii}]}

\label{firstpage}
\pagerange{\pageref{firstpage}--\pageref{lastpage}}
\maketitle

\begin{abstract}
A wide-frequency radio study of Active Galactic Nuclei (AGN) is crucial to evaluate the intervening radiative mechanisms responsible for the observed emission and relate them with the underlying accretion physics. We present wide frequency (5-45 GHz), high-sensitivity (few $\mathrm{{\mu}Jy\,beam^{-1}}$), (sub)-kpc JVLA observations of a sample of 30 nearby ($0.003\,\le\,z\,\le\,0.3$) AGN detected by {\it INTEGRAL/IBIS} at hard-X-ray. We find a high detection fraction of radio emission at all frequencies, i.e. $\ge$ 95 per cent at 5, 10 and 15 GHz and $\ge$ 80 per cent at 22 and 45 GHz. Two sources out of 30 remain undetected at our high sensitivities. The nuclear radio morphology is predominantly compact, sometimes accompanied by extended jet-like structures, or more complex features. The radio Spectral Energy Distributions (SEDs) of the radio cores appear either as single or as a broken power-law, a minority of them exhibits a peaked component. The spectral slopes are either flat/inverted or steep, up to a break/peak or over the whole range. The sample mean SED shows a flat slope up to 15 GHz which steepens between 15 and 22 GHz and becomes again flat above 22 GHz. Significant radio-X-ray correlations are observed at all frequencies.
About half of the sample features extended emission, clearly resolved by the JVLA, indicating low-power jets or large scale outflows. The unresolved cores, which often dominate the radio power, may be of jet, outflow, and/or coronal origin, depending on the observed frequency.
\end{abstract}

\begin{keywords}
galaxies: active --galaxies: nuclei -- galaxies: jets -- galaxies: Seyfert -- black hole physics -- radio continuum: galaxies
\end{keywords}

\section{Introduction}

The connection between the accretion and ejection physics in Active Galactic Nuclei (AGN) has been one of the hot topics of extragalactic astrophysics since the first discoveries of active galaxies. The existence of scaling relations involving the black-hole mass and fundamental quantities related to the host galaxy \citep[e.g.][]{KormendyRichstone1995,Magorrian1998,FerrareseMerritt2000,Gultekin2009} suggests that the AGN ejection activity and host galaxy evolution are entangled. Indeed, the mechanical kinetic energy ejected by the jet is believed to be one source of AGN feedback into the host galaxy \citep{Morganti2017}. A connection between the accretion and the ejection phenomena, with analogous properties to AGN, has been observed in other accreting systems like black-hole X-ray binaries, XRB \citep[e.g.][]{Falcke2004}, coronally-active stars \citep[e.g.][]{LaorBehar2008,RaginskiLaor2016}, young stellar objects \citep[e.g.][]{Price2003,FerreiraDeguiran2013}, cataclysmic variables \citep[e.g.][]{Kording2008}, ultra--luminous X-ray sources \citep[e.g.][]{Mezcua2013} and tidal disruption events \citep[e.g.][]{Zubovas2019}, suggesting a common underlying physics. 

Historically, AGN has been divided between Radio-Loud (RL) and Radio-Quiet (RQ)\footnote{The RL/RQ definition is based on the value of the ratio of optical-to-radio fluxes $R_O=\log{\frac{f{(4400{\mathring{A}}})}{f{(6\,cm)}}}$ \citep{Kellerman1989}. A similar definition has been proposed considering the radio-to-X-ray luminosities: $\mathrm{R_X=\log{({L_R(6\,cm)}/{L_{2-10\,keV}})}}$, with -4.5 as dividing threshold between RQ and RL \citep[e.g.][]{TerashimaWilson2003,Panessa2006}}. However, the radio-loudness definition strongly depends on the observed frequency, on the region size from where the radio flux is measured, and the boundaries between RL and RQ AGN may also depend on the accretion rate and luminosity of the source \citet{Panessa2007}. \citet{Padovani2016} proposed a more physically meaningful dichotomy between 'jetted' and 'non-jetted' sources, where 'jetted' sources are those exhibiting a clear signature of a relativistic jet. In this work we refer to RQ AGN in the sense of 'non-jetted' sources. There is emerging evidence that RQ AGN are not radio-silent, as in these objects we do observe radio emission in a variety of shapes, scales and strengths, with sources exhibiting outflowing and jet-like features, sub-relativistics, less powerful, less collimated and on smaller scales with respect to those of powerful RL AGN \citep[e.g.][]{Nagar2002}, see \citet{Panessa2019} for a review. 

\indent Previous works focusing on the cm spectral range of the RQ population at arcsec resolution generally reported moderately-to-high detection fractions, i.e. 70 - 90 per cent \citep[e.g.][]{vanderHulstCraneKeel981,Kukula1995,RoyColbertWilsonUlvestad1998,NagarWilsonMulchaeyGallimore1999,Thean2000}. More recently, \citet{Smith2016,Smith2020} found a high detection fraction ($\sim$96 per cent) for a sample of hard-X-ray selected AGN from Swift/BAT at 22 GHz with a $\sim$1 arcsec resolution. \\ 
\indent The morphology of the radio emission on arcsec scales is predominantly compact, either unresolved or marginally resolved. However, a number of sources exhibit extended structures, linear features, double/triple components, core plus jet like features as well as diffuse emission, on spatial scales ranging from $\sim$kpc \citep[e.g.][]{UlvestadWilsonSramek1981,Kukula1998,Leipski2006} down to few hundreds of pc \citep[e.g.][]{Kukula1995,SchmittUlvestadAntonucciKimney2001,Thean2001}. \\
\indent Previous studies of the spectral shape of RQ AGN were mainly performed at frequencies $\le$15 GHz, usually adopting a two-frequencies approach, which only gives a limited information on the overall shape due to the limited frequency coverage \citep[e.g.][]{Kellerman1989,Kukula1998,chen22}. Only few works have adopted a multi-frequency approach to study the spectral shape of RQ AGN, but still limited to the frequency range $\nu\,\le$15 GHz \citep[e.g.][]{AntonucciBarvainis1988,BarvainisLonsdaleAntonucci1996,Kukula1998,Barvainis2005}. These works have led to the finding that most sources show only steep spectra, however a number of them (roughly 40-50 per cent) exhibit flat/inverted spectra in the cm/sub-cm range \citep[e.g.][]{AntonucciBarvainis1988,Kukula1995,BarvainisLonsdaleAntonucci1996,Kukula1998,Barvainis2005,LaorBaldiBehar2019,Chiaraluce2019}. 

In a number of sources the flat/inverted slope is found to dominate only at higher frequencies ($\ge$ few GHz), i.e. an excess emission is found with respect to the extrapolation from lower frequencies. This trend has been first reported as a "high-frequency excess" up to 2 cm by \citet{AntonucciBarvainis1988,BarvainisLonsdaleAntonucci1996}, and then as a mm-excess, up to $\sim$ 100 GHz, by \citet{LaorBehar2008,Behar2015,Behar2018}. 
Several radiative processes have been invoked to explain the evidence of flat/inverted cm/sub-cm spectra in RQ AGN, both thermal and non-thermal, like optically-thick synchrotron emission from a compact source, either the base of the jet \citep[e.g.][]{FalckeMalkanBiermann1995,WilsonColbert1995,baldi21a,baldi22} or a magnetically-heated corona \citep[e.g.][]{LaorBehar2008,RaginskiLaor2016}, thermal free-free emission from nuclear sources at $\mathrm{T_B\,\sim\,10^6\,K}$ and $\mathrm{T_B\,\sim\,10^4\,K}$ emission from NLR or starburst \citep[e.g.][]{AntonucciBarvainis1988,BarvainisLonsdaleAntonucci1996,Padovani2011,Condon2013}. 

The evidence of steep spectra in RQ AGN has been explained invoking synchrotron or free-free emission in the optically-thin regime from disk or circumnuclear starburst regions, on host-galaxy scale but also down to few hundreds pc \citep[e.g.][]{BarvainisLonsdaleAntonucci1996,Padovani2011,Condon2013,Zakamska2016}; optically-thin synchrotron from AGN-driven outflows, usually on few kpc scale \citep[e.g.][]{Mundell2000,Jiang2010,Zakamska2014,LaorBaldiBehar2019}; or optically-thin synchrotron emission from plasma blobs in a sub-relativistic jet, usually associated to a linear morphology and on scales as small as hundreds of pc \citep[e.g.][]{BarvainisLonsdaleAntonucci1996,Kukula1999,LaorBaldiBehar2019}. 

As additional research channel, the study of correlations between the X-ray luminosity, considered a proxy for the accretion power, and the radio one, measuring the strength of ejection phenomena, has allowed to test different physical scenarios, such as radiatively efficient versus inefficient regimes \citep[e.g.][]{Merloni2003,Falcke2004,Coriat2011,DongWuCao2014,DongWu2015}, or the so-called coronal models \citep[e.g.][]{LaorBehar2008,RaginskiLaor2016}.

This work is a completion of the results presented in \citet{Chiaraluce2020}, which we hereafter refer to as Paper I. Here, we present the results from new Jansky Very Large Array (JVLA) multi-frequency observations and Spectral Energy Distributions (SEDs) for a sample of hard-X-ray selected AGN, in a wide range of frequencies, 5 - 45 GHz, with a high sensitivity, few ${\mu}Jy$/beam, and at a resolution of $\sim\,$ arcsec, i.e. spatial scales $\le$kpc.

In this paper we use the $\Lambda$-CDM with the following cosmological parameters: H$_o$=67.4 km s$^{-1}$ Mpc$^{-1}$, $\Omega_m$=0.315 and $\Omega_{\Lambda}$=0.685 \citep{planck2018}.


\section{The sample}

The starting point of our project is the volume limited sample of 89 hard-X-ray selected AGN observed by INTEGRAL/\textit{IBIS} and presented in \citet{Malizia2009}. We selected only the sources conveniently observable with the JVLA between 5 and 45 GHz, i.e. north of DEC -30 deg, and we considered only Seyfert galaxies, criteria which reduce the sample to 44 objects (hereafter 'parent' sample). The characteristics of this sample have been extensively presented in Paper I, and are here briefly retrieved.

\begin{figure*}
\scriptsize
\centering
\includegraphics[scale=0.28]{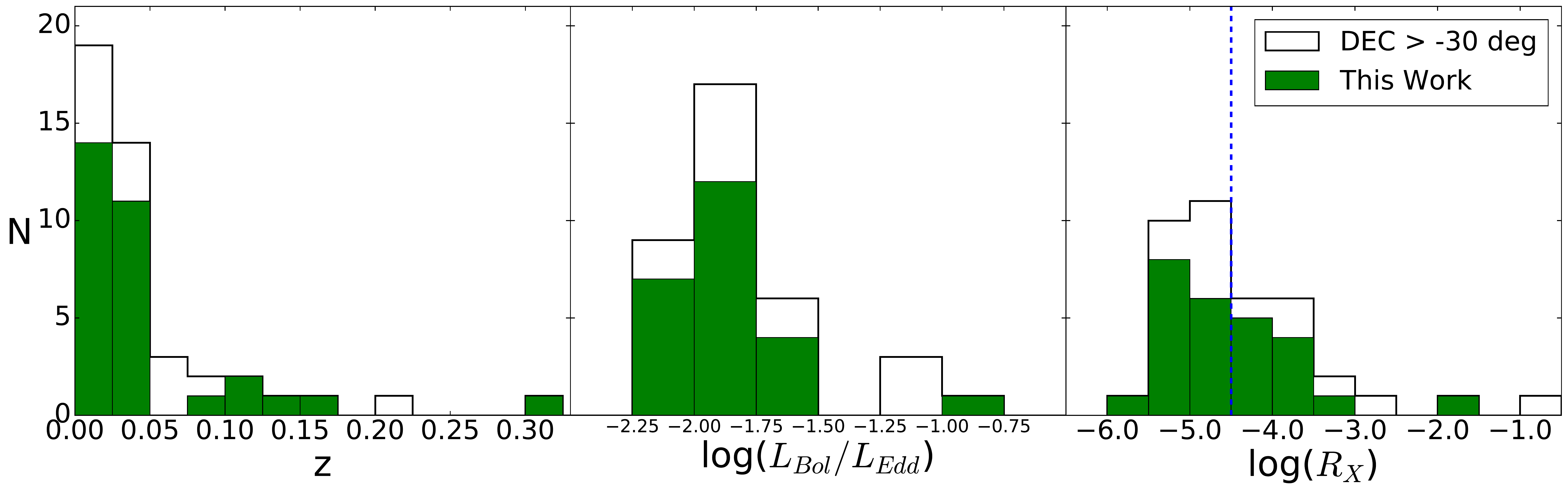}
\caption{Histogram showing the distribution of redshift \citep[from][left panel]{Malizia2009}. Eddington ratio ($L_{Bol}/L_{Edd}$. centre panel) and radio-loudness parameter \citep[from quantities tabulated in][right panel]{Panessa2015} for the parent sample (DEC $>$ -30 deg, semi-filled histogram) and for the sources considered in this work (green), i.e. having at least two SED data points. The blue, vertical dashed line is the R$_{X}\,=\,$-4.5 limit of \citet{TerashimaWilson2003}.}
\label{histogram_fundamental_quantities}
\end{figure*}

This hard-X-ray selected sample has several advantages: i) the hard-X-ray selection makes it relatively free of selection biases, e.g. absorption \citep[see discussion in][and Paper I]{HU01}; ii) it comprises moderate-to-high luminosity objects, with 41.5$\mathrm{\,\le\,\log{L_{2-10\,keV}}}$ $(erg\,s^{-1}$)$\,\le\,$44.5; iii) the accretion rates are relatively high: -2.5$\mathrm{\,\le\,\log{L_{Bol}}}$/${L_{Edd}}\,\le$-0.5 (central panel of Fig. \ref{histogram_fundamental_quantities}); iv) it covers a wide range of radio luminosities (Fig. \ref{histogram_HXAGN_SED_core_radio_luminosity}), with the bulk of the population made of RQ AGN (Fig. \ref{histogram_radio_loudness}).


\begin{table*}
\centering
\caption{List of sources together with observations information. \textit{Columns:} (1) Target name; (2) and (3) Right ascension and declination (J2000) from \citet{Malizia2009}; (4) Seyfert type; (5) Redshift; (6), (7), (8), (9) and (10), C, X, Ku, K and Q band projects, respectively. For the archival data we use the following notation: VLA project (month/year) [VLA configuration].}
\setlength\tabcolsep{2pt}
\scalebox{0.85}{
\begin{tabular}{lccccccccc}
\hline
 & & & & & \multicolumn{5}{c}{Observations} \\
\cline{6-10} \\
Target & RA & DEC & Seyfert & z & C & X & Ku & K & Q \\ 
  & (J2000) & (J2000) & Type &  & (5 GHz)  & (10 GHz) & (15 GHz) & (22 GHz) & (45 GHz) \\ 
 (1) & (2) & (3) & (4) & (5) & (6) & (7) & (8) & (9) & (10) \\ 
 \hline
 \hline
IGR~J00333+6122 & 00:33:18.41 & +61.27.43.10 & S1.5 & 0.105 & $\cdots$ & $\cdots$ & $\cdots$ & 18B-163 & 18B-163 \\ 
NGC~788 & 02:01:06.40 & $-$06:48:56.0 & S2 & 0.0136 & AW0126$\,$(02/85)$\,$[A] & AM384$\,$(12/92)$\,$[A]& $\cdots$ & 18B-163 & 18B-163 \\
NGC~1068 & 02:42:40.71 & $-$00:00:47.8 & S2 & 0.0038 & BG066$\,$(04/99)$\,$[B] & AZ0053$\,$(04/92)$\,$[C] & AW0085$\,$(01/84)$\,$[B] & 18B-163 & 18B-163 \\
QSO~B0241+62 & 02:44:40.71 & +62.28.06.5 & S1 & 0.044 & AM0124$\,$(06/86)$\,$[B/A] & AS644$\,$(11/98)$\,$[C] & AS0786$\,$(04/04)$\,$[C] & 18B-163 & 18B-163 \\ 
\hline
NGC~1142 & 02:55:12.19 & $-$00.11.02.3 & S2 & 0.0288 & 12A-066$\,$(06/12)$\,$[B] & \makecell{AP233$\,$(06/91)$\,$[A]\\AP212$\,$(04/92)$\,$[B]} & AJ105$\,$(05/84)$\,$[B] & 18B-163 & 18B-163 \\ 
\hline
B3~0309+411B & 03:13:01.96 & +41.20.01.2 & S1 & 0.136 & BT0007$\,$(05/94)$\,$[A/B] & BM0074$\,$(04/97)$\,$[B] & $\cdots$ & 18B-163 & 18B-163\\
NGC~1275 & 03:19:48.16 & +41.30.42.1 & S2 & 0.0175 &  $\cdots$ &  $\cdots$ &  $\cdots$ & 18B-163 & 18B-163\\
LEDA~168563 & 04:52::04.85 & +49.32.43.7 & S1 & 0.029 & $\cdots$ & $\cdots$ & $\cdots$ & 18B-163 & 18B-163 \\
4U~0517+17 & 05:10:45.51 & +16.29.55.8 & S1.5 & 0.0179 & $\cdots$ & $\cdots$ & $\cdots$ & 18B-163 & 18B-163 \\
MCG+08-11-11 & 05:54:53.61 & +46.26.21.6 & S1.5 & 0.0205 & ULVE$\,$(03/82)$\,$[A]& AB0973$\,$(11/04)$\,$[A] & AU0018$\,$(11/83)$\,$[A] & 18B-163 & 18B-163 \\
\hline
Mkn~3 & 06:15:36.36 & +71.02.15.1 & S2 & 0.0135 & \makecell{AW0258$\,$(08/90)$\,$[A]\\GK0011$\,$(02/94)$\,$[B]} & \makecell{AW0278$\,$(09/91)$\,$[A]\\BM0124$\,$(06/00)$\,$[B]} & AP0142$\,$(09/87)$\,$[A] & 18B-163 & 18B-163 \\
\hline
Mkn~6 & 06:52:12.25 & +74.25.37.5 & S1.5 & 0.0188 & \makecell{AB0740$\,$(07/95)$\,$[A]\\AB0740$\,$(11/95)$\,$[B]\\AB0740$\,$(05/95)$\,$[D]} & AW0278$\,$(09/91)$\,$[A] & AP0142$\,$(10/87)$\,$[A] & 18B-163 & 18B-163\\
\hline
NGC~4151 & 12:10:32.58 & +39:24:20.6 & S1.5 & 0.0033 & $\cdots$ & $\cdots$ & $\cdots$ & 18B-163 & 18B-163 \\
Mkn~50 & 12:23:24.14 & +02.40.44.8 & S1 & 0.0234 & 19A-018 & 19A-018 & 19A-018 & $\cdots$ & $\cdots$ \\
NGC~4388 & 12:25:46.75 & +12.39.43.5 & S2 & 0.0084 & 19A-018 & 19A-018 & 19A-018 & 18B-163 & 18B-163 \\
LEDA~170194 & 12:39:06.32 & -16.10.47.8 & S2 & 0.036 & 19A-018 & 19A-018 & 19A-018 & 20A-066 & 20A-066 \\
NGC~4593 & 12:39:39.42 & -05.20.39.3 & S1 & 0.009 & 19A-018 & 19A-018 & 19A-018 & 20A-066 & 20A-066 \\
IGR~J13091+1137 & 13:09:05.60 & +11.38.02.9 & S2 & 0.0251 & 19A-018 & 19A-018 & 19A-018 & 20A-066 & 20A-066 \\
NGC~5252 & 13:38:16.00 & +04.32.32.5 & S2 & 0.023 & 19A-018 & 19A-018 & 19A-018 & 18B-163 & 18B-163 \\ 
NGC~5506 & 14:13:14.87 & -03.12.26.9 & S1.9 & 0.0062 & 19A-018 & 19A-018 & 19A-018 & 20A-066 & 20A-066 \\
IGR~J16385-2057 & 16:38:30.91 & -20.55.24.6 & NLS1 & 0.0269 & $\cdots$ & $\cdots$ & $\cdots$ & 20A-066 & 20A-066 \\
\hline
IGR~J16426+6536 & 16:43:04.70 & +65:32:50.9 & NLS1 & 0.323 & $\cdots$ & $\cdots$ & $\cdots$ & \makecell{18B-163\\20A-066} & \makecell{18B-163\\20A-066} \\ 
\hline
IGR~J17513-2011 & 17:51:13.62 & -20.12.14.6 & S1.9 & 0.047 & 19A-018 & 19A-018 & 19A-018 & 20A-066 & 20A-066 \\
IGR~J18027-1455 & 18:02:45.50 & -14.54.32.0 & S1 & 0.035 & $\cdots$ & $\cdots$ & $\cdots$ & 20A-066 & 20A-066 \\
IGR~J18259-0706 & 18:25:57.58 & +07.10.22.8 & S1 & 0.037 & 19A-018 & 19A-018 & 19A-018 & 20A-066 & 20A-066 \\
2E~1853.7+1534 & 18:56:01.28 & +15.38.05.7 & S1 & 0.084 & $\cdots$ & $\cdots$ & $\cdots$ & 20A-066 & 20A-066 \\
IGR~J20186+4043 & 20:18:38.73 & +40.40.59.9 & S2 & 0.0144 & 19A-018 & 19A-018 & 19A-018 & 20A-066 & 20A-066 \\
4C~+74.26 & 20:42:37.30 & +75.08.02.4 & S1 & 0.104 & 19A-018 & 19A-018 & 19A-018 & 20A-066 & 20A-066 \\
IGR~J23308+7120 & 23:30:37.28 & +71.22.46.0 & S2 & 0.037 & 19A-018 & 19A-018 & 19A-018 & 20A-066 & 20A-066 \\ 
IGR~J23524+5842 & 23:52:22.11 & +58.45.30.7 & S2 & 0.164 & 19A-018 & 19A-018 & 19A-018 & 20A-066 & 20A-066 \\
\hline
\end{tabular}
}
\label{sources}
\end{table*}


\begin{table*}
\centering
\caption{Luminosities and radio loudness parameters: \textit{Columns:} (1) Target name; (2) 2-10 keV X-ray luminosity \citet{Malizia2009}; (3) 20-100 keV X-ray luminosity from \citet{Malizia2009}; (4) Bolometric luminosity; (5) Eddington ratio; (6) Optical radio-loudness parameter \citep{Kellerman1989}; (7) X-ray radio-loudness at 5 GHz; (8) X-ray radio-loudness at 22 GHz; (9) X-ray radio-loudness at 45 GHz.
}
\scalebox{0.85}{
\begin{tabular}{lccccccccc}
\hline
Name				&	log($L_{2-10~keV}$)	&	log($L_{20-100~keV}$)	& log($L_{Bol}$)  &	log($L_{Bol}/L_{Edd}$)	&	log($R_{O}$)	&	log($R_{X}^{5~GHz}$)	&	log($R_{X}^{22~GHz}$)	&	log($R_{X}^{45~GHz}$)	\\
                    &   (erg/s)             &   (erg/s)                 &   (erg/s) &&&&&\\
(1) & (2) & (3) & (4) & (5) & (6) & (7) & (8) & (9) \\ 
\hline
IGR~J00333+6122		&	44.19			&	44.44	&	44.80    &	-1.84	   &	1.90        & -4.42       & -4.64       & -4.71     \\								
NGC~788				&	42.86			&	43.29	&	43.55    &	-2.06	   &	-1.04       & -5.42       & -5.01       & -4.9    \\								
NGC~1068			&	42.95			&	43.35	&	43.36    &	-1.95	   &	0.04        & -3.92       & -4.11       & -4.06     \\									
QSO~B0241+62		&	43.87			&	44.41	&	44.39    &	-1.80	   &	0.70        & -3.06       & -1.91       & -1.60     \\									
NGC~1142			&	43.82			&	43.96	&	45.23    &	-2.27	   &	0.08        & -5.25       & -4.88       & -4.76     \\									
B3~0309+411B		&	45.04			&	44.99	&	$\cdots$ &	$\cdots$   &	3.28        & -3.00       & -1.96       & -1.67     \\						
NGC~1275			&	42.89			&	43.38	&	44.32    &	-2.28	   &	2.99        & -0.96       & -0.35       & -0.16     \\									
LEDA~168563			&	43.9			&	44.04	&	44.33    &	-1.77	   &	1.50        & -4.70       & -4.80       & -4.80     \\								
4U~0517+17			&	43.23			&	43.62	&	43.44    &	-1.66	   &	$\cdots$    & -5.27       & -4.75       & -4.40     \\								

MCG+08-11-11		&	43.69			&	43.74	&	43.90    &	-1.65	   &	1.27        & -4.25       & -4.29       & -3.98     \\								
Mkn~3				&	44.17			&	43.79	&	44.77    &	-2.03	   &	1.03        & -4.79       & -5.03       & -4.98     \\								
Mkn~6				&	43.27			&	43.57	&	44.27    &	-1.96	   &	1.17        & -3.93       & -3.40       & -3.31     \\								
NGC~4151			&	43.05			&	43.16	&	43.89    &	-1.91	   &	0.60        & -5.41       & -4.84       & -4.73     \\									
Mkn~50				&	43.05			&	43.2	&	43.66    &	-1.84	   &	-1.02       & -5.53       & $\cdots$    & $\cdots$  \\								
NGC~4388			&	43				&	43.54	&	43.16    &	-1.80	   &	-0.80       & -4.94       & -5.05       & -4.83     \\									
LEDA~170194			&	43.48			&	44.22	&	44.76    &	-2.24	   &	0.30        & -4.42       & -3.67       & -3.20     \\								
NGC~4593			&	42.8			&	43.09	&	43.12    &	-1.72	   &	-1.86       & -5.63       & -4.95       & -4.49     \\									
IGR~J13091+1137		&	43.44			&	43.77	&	44.44    &	-2.16	   &	-0.90       & -5.49       & -5.17       & -5.08     \\								
NGC~5252			&	43.72			&	43.71	&	44.89    &	-2.21	   &	0.60        & -4.71       & -4.39       & -4.60     \\									
NGC~5506			&	42.83			&	42.85	&	42.96    &	-1.79	   &	0.90        & -3.91       & -3.85       & -3.77     \\									
IGR~J16385-2057		&	43.02			&	43.39	&	43.26    &	-1.67	   &	$\cdots$    & $\cdots$    & -4.37       & -4.32     \\
IGR~J16426+6536		&	$\cdots$		&	45.97	&  $\cdots$  &	$\cdots$   &	$\cdots$    & $\cdots$    & $\cdots$    & $\cdots$  \\			
IGR~J17513-2011		&	43.4			&	44.1	&	43.09    &	-1.01	   &	$\cdots$    & -4.75       & -4.29       & -3.89     \\
IGR~J18027-1455		&	43.23			&	44.04	&	43.83    &	-1.83	   &	$\cdots$    & $\cdots$    & $<$-5.46    & $<$-4.97  \\						
IGR~J18259-0706		&	43.19			&	43.66	&	$\cdots$ &	$\cdots$   &	$\cdots$    & -4.99       & -4.65       & -4.37     \\					
2E~1853.7+1534		&	44.25			&	44.56	&	44.55    &	-1.70	   &	$\cdots$    & $\cdots$    & -5.64       & $<$-5.17  \\							
IGR~J20186+4043		&	42.57			&	42.98	&	$\cdots$ &	$\cdots$   &	-0.67       & -4.72       & -4.62       & -4.60     \\					
4C~+74.26			&	44.75			&	45.04	&	45.76    &	-1.96	   &	1.92        & -3.20       & -2.54       & -2.45     \\								
IGR~J23308+7120		&	42.8			&	43.55	&	$\cdots$ &	$\cdots$   &	-0.44       & -5.16       & $<$-5.11    & $<$-4.28  \\					
IGR~J23524+5842		&	44.41			&	44.85	&	$\cdots$ &	$\cdots$   &	1.12        & -5.04       & -5.22       & $<$-4.67  \\
\hline				
\end{tabular}
}
\label{quantities}
\end{table*}


\section{Radio observations and data reduction}

In the present work we couple our high-frequencies radio data at 22 and 45 GHz (JVLA-C configuration, project 18B-163, $\sim$1 and 0.5 \arcs resolution, respectively) already published in Paper I, to lower frequencies (5, 10 and 15 GHz, i.e., C, X and Ku bands respectively) proprietary data obtained in JVLA-B configuration (project: 19A-018). The more extended configuration at lower frequencies guarantees us a matched resolution with high frequencies data, a crucial point in order to build physically meaningful SEDs: at 5, 10 and 15 GHz the expected resolution is $\sim$1, 0.6 and 0.4\arcs. Moreover, we complemented the high-frequency observations at 22 and 45 GHz with new ones obtained in the same configuration (project 20A-066). 

Finally, we searched in the VLA archive low-frequency data for the objects for which our own proprietary data were not available (i.e. NGC~788, NGC~1068, QSO~B0241+62, NGC~1142, B3~0309+411B, MCG~+08-11-11, Mkn~3 and Mkn~6). When available, we selected observations in the same configurations (VLA-B), when not we considered other available configurations (mostly A), and in order to match the resolution a tapering has been applied (next Section).

The above strategy translates into 30 objects out of 44 with at least two SED data points. In Table \ref{sources} we summarise the characteristics of the sample together with information on the observations and the corresponding projects: 9 sources have only two SED data points; one source (Mkn~50) has three SED data points; two sources have four data points (NGC~788 and B3~0309+411B), the majority of sources (18 out of 30) have 5 SED data points. 

In Fig. \ref{histogram_fundamental_quantities} we plot a histogram of the distribution of the relevant physical quantities of sources in this work, i.e redshift, Eddington ratio and X-ray radio-loudness ($\mathrm{R_X=\log{({L_R(6\,cm)}/{L_{2-10\,keV}})}}$), as compared to the whole sample of 44 source north DEC $\ge$ -30 deg. A Kolmogorov-Smirnov test comparing the two distributions results, in all cases, in a P$\sim$0.99, so the null hypothesis that the two samples are drawn from the same parent population cannot be rejected. This would suggest that the physical conclusions which can be drawn from the sources considered in this work are likely representative of the whole sample. 

The project 19A-018 has been observed in the period February - May 2019, the project 20A-066 in the period February - May 2020. The characteristics of observations for the high-frequency project 20A-066 are the same of 18B-163, presented in Section 3 of Paper I, and we refer to that section for the details. Sources in 19A-018 were observed at 5, 10 and 15 GHz for a minimum of 1 min to a maximum of 5 min.
We applied phase calibration, bracketing targets with suitable sources within a radius of 10 deg.
The scans of the absolute flux density scale calibrator has been performed once per scheduling block, typically at the beginning or end of it, and, at a given frequency, it is observed for no less than 2min30s.

The data calibration and reduction procedure for the proprietary JVLA data has been performed with the calibration pipeline within the Common Astronomy Software Application \citep[\textsc{casa} 5.4.1 version\footnote{\url{https://casa.nrao.edu}},][]{McMullin+07}. 
After calibration, the plots were inspected for residual RFI. For the image reconstruction, the \textsc{tclean} task in \textsc{casa} has been used, applying the \cite{1974A&AS...15..417H} deconvolver. 
We produced full resolution maps considering as initial weighting the \citet{Briggs1995} one with robustness parameter equal to 0.5, which ensures a balance between resolution and sensitivity. However, in order to have X and Ku maps at a matched resolution with respect to C band, we produced naturally-weighted tapered maps, in order to give more weight to the short baselines and have an approximately equal UV coverage.

The typical RMS achieved is $\sim10-40\,$\mujybm, in agreement with proprietary post-upgrade JVLA data (see Table 3 in Supplementary material).
For the archival VLA data, the reduction has been performed with the Astronomical Image Processing System \textsc{AIPS} in the 31DEC20 version\footnote{\url{http://www.aips.nrao.edu}}, following standard procedures. The imaging procedure has been conducted via the task \textsc{imagr}.
In all cases, self--calibration has been performed for all the sources strong enough to provide enough flux density for the model.

For both the proprietary JVLA data and the archival VLA ones, the image analysis has been performed via the \textsc{casa} task \textsc{imfit}. Based on the source morphology, single/multiple Gaussian fit on the image plane has been performed.
A comparison of the flux densities of calibrators for the proprietary JVLA data with tabulated values in \citet{PerleyButler2017} led to the estimation of a $\sim$5 per cent error in the flux calibration; for the archival data a more conservative 10 per cent has been adopted. The uncertainties in the final flux density measurements are affected by fitting errors from \textsc{imfit}, and flux calibration error, which are added in quadrature and adopted as the error measurements. In the case of components with a resolved morphology, the error associated with the flux determination is given by the formula ${\sigma_S}=\sqrt{N\times{(rms)^2}+({A_{s}}{\times}S)^2}$, where N is the number of beam areas covered by a source of flux density S, and $A_{S}$ takes into account the  uncertainty in the absolute flux density scale \citep[0.05$-$0.1, see][]{HU01}.

In Table \ref{sources} we report the source coordinates taken from \citet{Malizia2009}. In a number of sources the above positions display a significant offset from the derived radio ones (the latter reported in Table 3 in Supplementary material). For this reason, we performed a careful cross-check with all available positions (from optical, X-ray, IR, etc observations) in literature\footnote{With the NASA NED and SIMBAD databases.} for each source, in order to determine the likely core component. In the maps, the white (or black) crosses represent the original optical coordinates as in \citet{Malizia2009}, as quoted in Table \ref{sources}, that were adopted during the scheduling of our observations and therefore are their phase centres. In Table 2 in Supplementary material we report the integrated and peak radio luminosities of the core component at all frequencies and in Table 3 in Supplementary material we report the derived parameters from the imaging analysis. 
The complete set of maps for the entire sample is available on the project's website\footnote{\url{https://sites.google.com/inaf.it/hxagn/radio-images}}, while a representative image for each source, together with its SED, is reported in Figure 2 in Supplementary material.
In Table \ref{quantities} we report the X-ray (2-10 keV and 10-100 keV) and bolometric luminosities and the corresponding Eddington ratios (as detailed in Paper I). The optical radio-loudness parameter \citep{Kellerman1989} and the X-ray radio-loudness parameter \citep{TerashimaWilson2003} is calculated using the radio luminosities at 5, 22 and 45 GHz.


\section{Results}

\subsection{Detection rates, radio luminosities and radio loudness}

We find high detection rates of radio emission at all frequencies, with detection fractions which decrease with frequency, i.e. 21/21, 21/21, 19/19, 26/29 and 24/29 at 5, 10, 15, 22 and 45 GHz, respectively, which translate into rates of 96$\pm$8, 96$\pm$8, 95$\pm$8, 87$\pm$12 and 81$\pm$14, respectively\footnote{Due to the limited statistics of the sample, in order to estimate the detection fractions and associated uncertainties we used the Laplace point estimate formula (Laplace 1812) and the Adjusted Wald method \citep[e.g.][]{SauroLewis2005}.}.


\begin{figure*}
\centering
\includegraphics[scale=0.35]{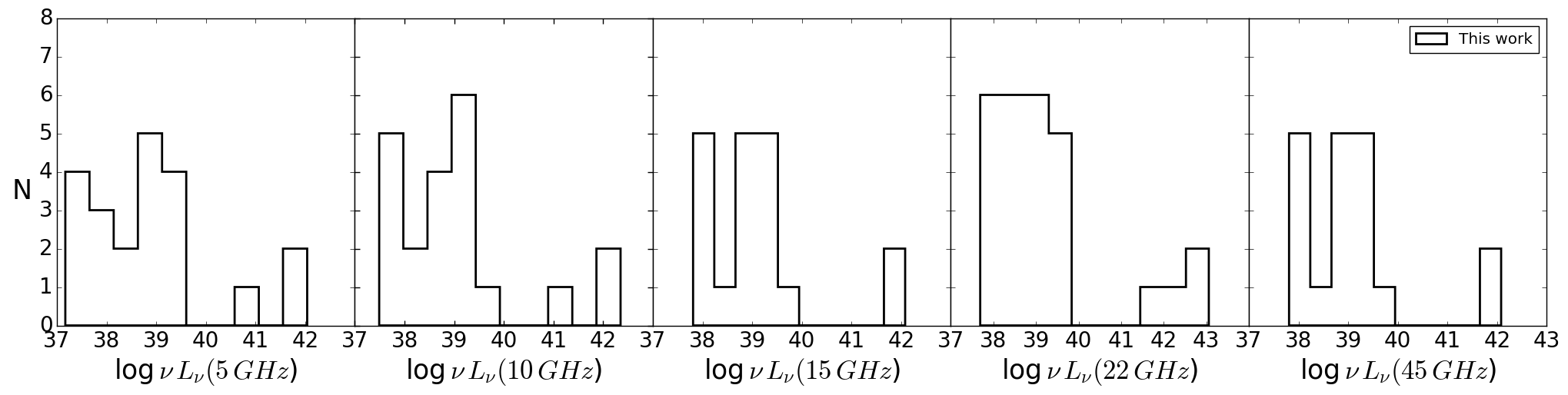}
\caption{Histogram showing the distribution of core radio luminosities $\mathrm{\log{{\nu}L_{\nu}}}$ at the different frequencies for sources in the sample.}
\label{histogram_HXAGN_SED_core_radio_luminosity}
\end{figure*}


In Fig. \ref{histogram_HXAGN_SED_core_radio_luminosity} we show the distribution of core radio luminosities for the sources in the sample at different frequencies. The bulk of sources has core radio luminosities in the range 37$\mathrm{\,\le\,\log{\,\nu\,L_{\nu}(erg\,s^{-1}})\,\le\,}$40, 
well below the values observed in radio galaxies (i.e., $\mathrm{\log{\,\nu\,L_{\nu}(erg\,s^{-1}})>}$42, \citet{fr95}).

\begin{figure*}
\scriptsize
\centering
\includegraphics[scale=0.225]{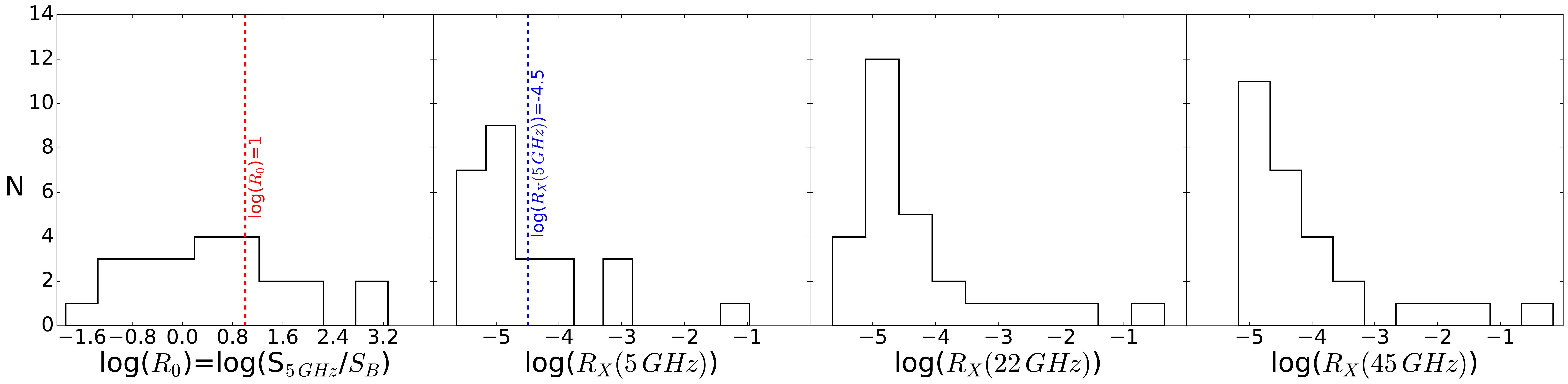}
\caption{Histogram showing the distribution of radio loudness for the sources in the sample based on different definitions, i.e. \citet{Kellerman1989} (first panel, from left, 9/30 classify as RL); \citet{TerashimaWilson2003} at 5 GHz (second panel, 10/30 classify as RL), 22 GHz (third panel) and at 45 GHz (fourth panel). The vertical lines denote the RQ/RL limit depending on the definition.}
\label{histogram_radio_loudness}
\end{figure*}

In Fig. \ref{histogram_radio_loudness} we show the distribution of radio-loudness parameters of the core component at different frequencies. The fraction of radio sources which can be classified as RL is 30 and 33 per cent considering the classical radio loudness parameter and the X-ray one at 5 GHz, respectively \citep{TerashimaWilson2003}. However, given the low radio luminosities and the caveats associated to the definition of a boundary in the radio-loudness parameter, we consider as RQ AGN the majority of the sources in our sample, except for the three most powerful radio emitters already mentioned in Paper I, i.e. QSO~B0241+62, B3~0309+411B and NGC~1275, and the giant radio galaxy 4C~+74.26 \citep[e.g.][and references therein]{Bruni2020}, resulting in $\sim$ 13\% of RL AGN in the sample. This fraction of RL AGN is consistent with what already observed so far in hard X-ray selected samples \citep[e.g.][and references therein]{Bassani16}.


\subsection{Undetected sources}

Two sources out of 30 were not detected by our observations, i.e. IGR~J16426+6536 and IGR~J18027-1455, and for both only observations at 22 and 45 GHz were available. For both we give 3$\times\,\sigma$ upper limits of $\sim\,$0.1 and $\sim\,$0.15 \mjybm{}{} at 22 and 45 GHz, respectively, which result, in the case of IGR~J18027-1455, in upper limits on the radio luminosities of $\mathrm{\log{L_R\,(erg\,s^{-1})}\,\sim}$ 37.8 and 38.3 (upper limits on the radio powers of $\mathrm{L_{\nu}(W\,Hz^{-1})\,\sim\,3\times10^{20}}$ and $\mathrm{L_{\nu}(W\,Hz^{-1})\,\sim\, 4.5\times10^{20}}$ at 22 and 45 GHz, respectively). The derived upper limits on the X-ray radio-loudness parameter are of -5.4 and -4.9 at the two frequencies, respectively, placing these sources at the very-low-power end of the radio-loudness distribution. The source IGR~J16426+6536 benefits from observations in two epochs, one in November 2018 and presented in Paper I, and one in June 2020, presented here. In both cases the source is undetected at a similar flux limit (see Table 3 in Supplementary material), variability at 17 months scales can therefore be ruled out. Deeper observations, achievable with longer integration times, would be required to confirm the radio-silent nature of these sources.


\onecolumn
\begin{landscape}

\begin{figure*}
\centering
\includegraphics[trim=+1cm 0 0 -1cm,scale=0.6]{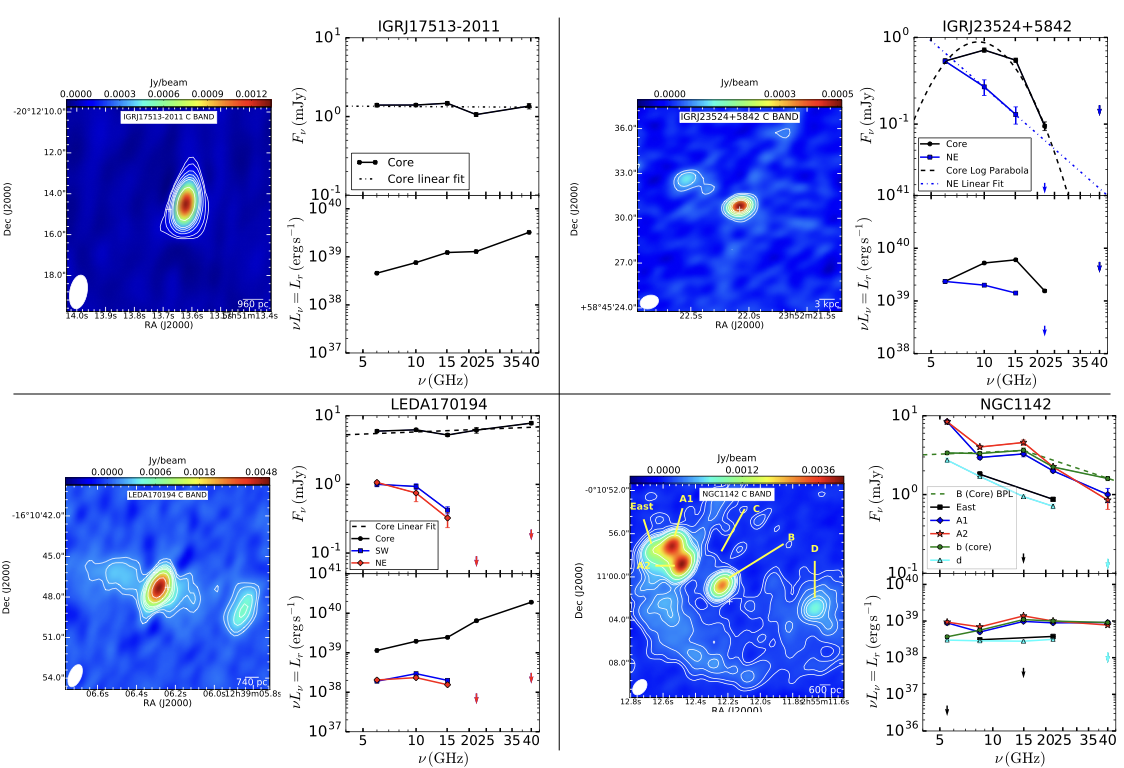}
\caption{Radio maps of sources considered as an example for the four morphological types adopted in Section 4.3: top-left, IGR~J17513-2011 (morphological class A); top-right, IGR~J23524+5842 (morphological class B); bottom-left, LEDA~170194 (morphological class C); bottom-right, NGC~1142 (morphological class D). For space reasons, only the 5-GHz (C-band) maps are shown. For each source we also show the corresponding SED, in which we indicate the core and all the other emitting components (if present). The SED and morphology for the remaining sources are shown in Figure 2 in Supplementary material.}
\label{fig:4classes}
\end{figure*}

\end{landscape}
\twocolumn


\subsection{Morphology}

We divided the morphology of the sources in the sample into four main classes, adapting the classification in \citet{baldi21b}, considering both full array and tapered radio maps. In Table \ref{spectral_indices}, we report the morphological classification (column 2) for each source and in Fig. \ref{fig:4classes} we report an examples of each of the four morphological classes:
\begin{itemize}
    \item \textbf{core or core+jet/lobe (A)}: sources with a single core component, either resolved or marginally resolved, sometimes with a small-scale protrusion (either one-sided or two-sided); 17/30 sources belong to this class (e.g., IGRJ~17513-2011).
     \item \textbf{one-sided jet (B)}: sources with an extended asymmetric jet, which is possibly resolved into several emitting components at higher resolutions; 7/30 sources are in this class (e.g., IGRJ~23524+5842).
    \item \textbf{triple (C)} sources exhibiting a triple morphology, interpreted as the radio core plus symmetric jet/lobes (one source: LEDA~170194). \item \textbf{jet+complex (D)}: sources with a complex structure, with multiple components, either compact, diffuse and/or extended; three sources fall in this classification, i.e. NGC~1068, NGC~1142 and NGC~5506. Note that the D classification corresponds to the E classification in \citet{baldi21b}.
\end{itemize}
    
\indent The morphological study for the sources across the cm-wave range reveals a prevalence of compact components, either unresolved or marginally-resolved, with nearly a half of the sample sources exhibiting jet-like structures, as well as triple and more complex ones that extend up to kpc scales. We note that the morphological classification depends on the distance of the source, i.e., for high z sources structures below the kpc scales are not resolved. However the majority of the sample sources are located in the nearby Universe, allowing us to map projected linear sizes of hundreds of pc (e.g. the nearest source is NGC~4593, $\sim$ 190 pc). On the other hand, high z objects (z$>$ 0.1) in our sample are mostly radio galaxies (e.g. 3C+74.26, $\sim$ 2 kpc) with a known morphology.
The classification also depends on the selected frequency. As an example, MCG+08-11-11 appears as a core plus a one sided jet-like feature in our maps at matched resolution of $\sim$1 arcsec, however, in the full resolution image, the core is resolved into 3 components, the central one being the core \citep[e.g.][]{UlvestadWilson1986}.
Indeed, sources with extended emission are best detected at low frequencies for a combination of effects. The surface brightness is enhanced by the steep spectrum and for a given JVLA configuration, the 
beam is larger, increasing the sensitivity to low surface brightness emission.


\subsection{Radio spectral energy distributions}

\begin{table*}\footnotesize
\centering
\caption{The two-frequency spectral indices (and associated uncertainties) for the sources and the multiple components, when present. The frequency pairs considered are 5-10 GHz, 10-15 GHz, 15-22 GHz, 22-45 GHz; for NGC~788 and B3~0309+411B we report the 10-22 GHz spectral index, due to the lack of 15 GHz data. We also indicate the morphological classification based on the four classes.}
\footnotesize
\setlength\tabcolsep{4pt}
\begin{tabular}{lcccccccccccccc}
\hline
Name & Radio & Component &  $\alpha_{5-10}$ & $\alpha_{10-15}$ &   $\alpha_{15-22}$ & $\alpha_{22-45}$ &  $\alpha_{10-22}$ \\
& morph & &   &   &  &   & \\
\hline
IGR~J00333+6122 & A & core & $\cdots$ & & & 1.2 $\pm$ 0.4 & \\
NGC~788 & A & core & 0.8 $\pm$ 0.2 & $\cdots$ & $\cdots$ & 0.4 $\pm$ 0.3 & -0.05$\pm$0.20\\
NGC~1068 &   & a & 0.65 $\pm$ 0.13 & 1.64 $\pm$ 0.13 & 0.69 $\pm$ 0.17 & 0.9 $\pm$ 0.1 & \\
		& D & b(core) & 0.49 $\pm$ 0.13 & 1.59 $\pm$ 0.13 & 1.1 $\pm$ 0.2 & 0.8 $\pm$ 0.1 & \\
        &   & d & $\cdots$ & $\cdots$ & 0.62 $\pm$ 0.18 & 0.47 $\pm$ 0.01 & \\
        &   & sw & $\cdots$ & $\cdots$ & -0.24 $\pm$ 0.19 & 0.14 $\pm$ 0.10 & \\
QSO~B0241+62 & B & core & -0.86 $\pm$ 0.13 & -1.34 $\pm$ 0.12 & 0.252 $\pm$ 0.180 & 0.3 $\pm$ 0.1 &\\
            &   & blob & 0.57 $\pm$ 0.13 & 2.4 $\pm$ 0.2 & $\cdots$ & $\cdots$ &  \\
NGC~1142 & D & east & $\cdots$ & $\cdots$  & $\cdots$ & $\cdots$ & \\
		&   & a1 & 2.45 $\pm$ 0.18 & -0.18 $\pm$ 0.18 & 1.28 $\pm$ 0.25 & 0.9 $\pm$ 0.3 & \\
		& 	& a2 & 1.72 $\pm$ 0.17 & -0.24 $\pm$ 0.16 & 1.83 $\pm$ 0.23 & 1.4 $\pm$ 0.1 & \\
		&	& b(core) & 0.04 $\pm$ 0.17 & -0.17 $\pm$ 0.15 & 1.26 $\pm$ 0.21 & 0.5 $\pm$ 0.1 & \\
		& 	& d & 1.11 $\pm$ 0.18 & 1.03 $\pm$ 0.12 & 0.73 $\pm$ 0.22 & $\cdots$ & \\
B3~0309+411B & B & core & -0.38 $\pm$ 0.13 &  $\cdots$ & $\cdots$ & 0.06 $\pm$ 0.10 & -0.7$\pm$0.1\\
			&	& nw & 0.11 $\pm$ 0.13 & $\cdots$ & $\cdots$ & $\cdots$ & \\
NGC~1275 & A & core & $\cdots$ & $\cdots$ & $\cdots$ & 0.4 $\pm$ 0.1 & \\
LEDA~168563 & A & core & $\cdots$ & $\cdots$ & $\cdots$ & 1.07 $\pm$ 0.12 & \\
4U~0517+17 & A & core & $\cdots$ & $\cdots$ & $\cdots$ & -0.09 $\pm$ 0.11 & \\
MCG+08-11-11 & B/C & core & 1.07 $\pm$ 0.13 & 1.1 $\pm$ 0.1 & 0.14 $\pm$ 0.19 & 0.7 $\pm$ 0.1 & \\
			 & 	   & jet  & 2.23 $\pm$ 0.13  & 2.5 $\pm$ 0.5  & -3.2 $\pm$ 0.9  & 1.1 $\pm$ 0.4 & \\
Mkn~3 &   & west       & 1.31 $\pm$ 0.13 & 0.77 $\pm$ 0.12 & 1.8 $\pm$ 0.3 & 1.2 $\pm$ 0.1 & \\
	 & B & east(core) & 1.41 $\pm$ 0.13 & 0.35 $\pm$ 0.10  & 1.3 $\pm$ 0.1 & 0.87 $\pm$ 0.09 & \\
Mkn~6 & 		 & N       & 0.98 $\pm$ 0.07 & 1.13 $\pm$ 0.10 & $\cdots$ & $\cdots$ & \\
     & B/C/D & S(core) & 0.91 $\pm$ 0.07 & 1.0 $\pm$ 0.1 & $\cdots$ & $\cdots$ & \\
	 &	     & Ext-EW  & 2.6 $\pm$ 0.1   & $\cdots$ & $\cdots$ & $\cdots$ & \\
NGC~4151 & A & core & $\cdots$ & $\cdots$ & $\cdots$ & 0.7 $\pm$ 0.1 & \\
Mkn~50 & A & core & -0.06 $\pm$ 0.20 & 0.03 $\pm$ 0.23 & $\cdots$ & $\cdots$ & \\
NGC~4388 & B(core) & NE & 0.9 $\pm$ 0.1 & 0.8 $\pm$ 0.2 & 1.08 $\pm$ 0.20 & 0.27 $\pm$ 0.11 & \\
        &         & SW & 0.8 $\pm$ 0.1  & 0.47 $\pm$ 0.18 & 1.45 $\pm$ 0.22 & 1.6 $\pm$ 0.3 & \\
		&         & North & 1.11 $\pm$ 0.13 & 5.4 $\pm$ 0.1 & $\cdots$ & $\cdots$ & \\
LEDA~170194 & C & Core & -0.09 $\pm$ 0.14 & 0.42 $\pm$ 0.17 & -0.21 $\pm$ 0.14 & -0.8 $\pm$ 0.4 & \\
		   &   & SW   & 0.14 $\pm$ 0.24 & 1.9 $\pm$ 0.4 & $\cdots$ & $\cdots$ & \\
		   &   & NE   & 0.7 $\pm$ 0.5 & 2.0 $\pm$ 0.9 & $\cdots$ & $\cdots$ & \\
NGC~4593 & A & core & -1.01 $\pm$ 0.15 & -0.23 $\pm$ 0.16 & 0.77 $\pm$ 0.18 & -0.54 $\pm$ 0.12 & \\
IGR~J13091+1137 & A & core & -0.57 $\pm$ 0.16 & -0.51 $\pm$ 0.19 & 2.6 $\pm$ 0.2 & 0.7 $\pm$ 0.4 & \\
NGC~5252 & A & core & 0.04 $\pm$ 0.11 & 0.61 $\pm$ 0.17 & 0.28 $\pm$ 0.20 & 1.7 $\pm$ 0.1 & \\
NGC~5506 & D & core & 0.95 $\pm$ 0.15 & 0.78 $\pm$ 0.11 & 1.1 $\pm$ 0.2 & 0.76 $\pm$ 0.11 & \\
IGR~J16385-2057 & A & core & $\cdots$ & $\cdots$ & $\cdots$ & 0.74 $\pm$ 0.25 & \\
IGR~J17513-2011 & A & core & -0.004 $\pm$ 0.140 & -0.13 $\pm$ 0.18 & 0.9 $\pm$ 0.2 & -0.40 $\pm$ 0.14 & \\
IGR~J18259-0706 & A & core & 0.86 $\pm$ 0.14 & 0.8 $\pm$ 0.2 & -0.6 $\pm$ 0.4 & 0.07 $\pm$ 0.30 & \\
2E~1853.7+1534 & A & core & $\cdots$ & $\cdots$ & $\cdots$ & $\cdots$ & \\
IGR~J20186+4043 & A & core & 0.87 $\pm$ 0.14 & 0.44 $\pm$ 0.17 & 1.3 $\pm$ 0.2 & 0.9 $\pm$ 0.3 & \\
4C~+74.26 & A & core & -0.15 $\pm$ 0.09 & 0.003 $\pm$ 0.200 & 0.24 $\pm$ 0.20 & 0.7 $\pm$ 0.1 & \\
IGR~J23308+7120 & A & core & 0.64 $\pm$ 0.30 & -0.4 $\pm$ 0.4 & $\cdots$ & $\cdots$ & \\
IGR~J23524+5842 & B & core & -0.6 $\pm$ 0.2 & 0.65 $\pm$ 0.20 & 4.63 $\pm$ 0.34 & $\cdots$ & \\
 &   & NE   & 1.33 $\pm$ 0.40 & 1.8 $\pm$ 0.7 & $\cdots$ & $\cdots$ & \\
\hline
\end{tabular}
\label{spectral_indices}
\end{table*}


\begin{figure*}
\scriptsize
\centering
\includegraphics[scale=0.5]{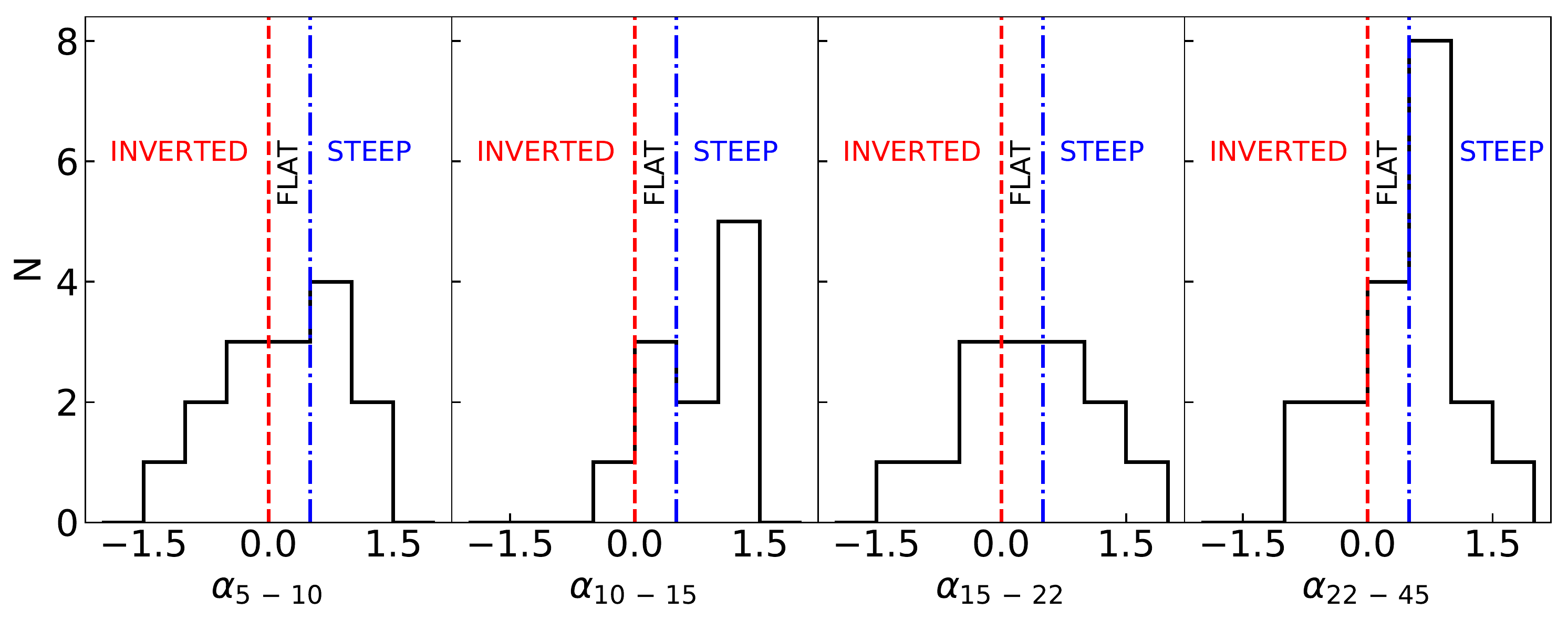}
\caption{Histogram showing the distribution of two-bands spectral indices ($S_{\nu}^I\,\propto\,{\nu}^{-\alpha}$) for the core components of sources in this work; i.e. $\alpha_{5\,-\,10}$, $\alpha_{10\,-\,15}$, $\alpha_{15\,-\,22}$ and $\alpha_{22\,-\,45}$ from left to right. The region with $\alpha\,<\,0$ and delimited by the red dashed line represents the 'inverted' regime; the region with $0<\,\alpha\,<\,0.5$ between the red dashed and the blue dot-dashed lines represents the 'flat' regime; the region with with $\alpha\,>\,0.5$ and delimited by the blue dot-dashed line represents the 'steep' regime, according to the adopted definitions.}
\label{histogram_spectral_indices}
\end{figure*}


With the present observations, spanning the 5 - 45 GHz interval, we can build spectral energy distributions in the cm/sub-cm range at a matched resolution ($\sim$\ arcsec) and sensitivity (a few \mujybm) for a statistically relevant sample. In this way it is possible to characterise the wide-frequency emission for the emitting components, both nuclear and extra-nuclear, and interpret them in light of models invoked for emission in RQ objects. \\
The definition of the spectral index used in this work is $S_{\nu}^I\,\propto\,\nu^{-\alpha}$, where $S_{\nu}^I$ is the integrated flux density. 
The uncertainties associated with the spectral indices has been estimated as $\sqrt{{{(\sigma_{f_1}/S_{f_1})}^2}+{{(\sigma_{f_2}/S_{f_2})^2}}}/ln({f_2/f_1})$, where $\sigma_{f_{1,2}}$ and $S_{f_{1,2}}$ are the uncertainties on the flux density and the flux density at the two frequencies \citep{HU01}, which are the central frequencies of the K and Q bands (therefore the flux densities are the mean across the bandwidth). We define a steep spectrum as having $\alpha$ $\ge$ 0.5, and a flat one as having
$\alpha$ $<$ 0.5, as in \citep{PG13}. We define a spectrum as inverted if $\alpha$ $<$ 0.

In order to characterise the spectral shape of the core components, we adopted the classification presented in \citet{BarvainisLonsdaleAntonucci1996}, (see also \citet{scheuer68}), i.e. single power-law (S) spectra, for which a simple power-law fit is sufficient, concave (C+) and convex (C-) spectra, for which the spectrum can be fitted by a quadratic polynomial, and complex spectra (CPX), which do not fit in none of the above cases. Based on the evidence in the sample, we add an extra class, the spectra for which a broken power-law (BPL) fit represents well the data. \\
\indent In Table \ref{SED_fitting} we report the results of the SED fitting, first for the broken power-law sources, then for the convex sources and finally for the single power-law ones. In Fig. 2 in Supplementary material we show the spectral flux and energy distributions of sources across the frequency range covered by our observations (sources are ordered as in Table \ref{SED_fitting}), while in Fig. \ref{histogram_spectral_indices} we show the distribution of two-frequency spectral indices for the core components of sources in the sample for the frequency pairs 5-10 GHz, 10-15 GHz, 15-22 GHz and 22-45 GHz. 
The relative fraction of sources in the different classes is the following: 67\% of single power-law spectra (e.g. left panels of Fig. \ref{fig:4classes}),  14\% of convex spectra (e.g. top-right panel of Fig. \ref{fig:4classes}), 19\% of broken power-law spectra (e.g. bottom-left panel of Fig. \ref{fig:4classes}),  none of concave nor complex spectra. However, cases which have been classified as convex, i.e. NGC~4593 and IGR~J13091+1137, could be classified in the complex class as well (see next sections). In this classifications we excluded the sources with less than 2 frequency points (6/30 with two data points, 1/30 with one and 2/30 undetected). \\
\indent The extra-nuclear components, when present, generally exhibit steep spectra. 


\begin{table*}\footnotesize
\centering
\caption{Results of the SED fitting for the sources in the sample. \textit{Columns:} (1) Target name; (2) Component; (3) classification of core SED; (4) Spectral index from linear fit to all points; (5) Break frequency (in GHz) in case of broken power-law fit or peak in case of log-parabola fit; (6) Spectral index at $\nu\le\nu_{B}$ (before break or peak); (7) Spectral index at $\nu\ge\nu_{B}$ (after break or peak). We show first the sources for which the core can be fitted by a broken power law, then those for which we performed a log-parabolic fit, finally the sources for which the core can be fitted by a single power-law. The sources for which only two data points or less were available (6/30) are omitted.}
\footnotesize
\begin{tabular}{lcccccc}
\hline
Target & Comp & SED class & $\alpha_{All}$ & $\nu_{Break/Peak}$ & $\alpha(\nu\le\nu_{Break/Peak})$ & $\alpha(\nu\ge\nu_{Break/Peak})$ \\ 
 (1) & (2) & (3) & (4) & (5) & (6) & (7)\\ 
 \hline
 \hline
NGC~5252 & Core & BPL & - & $\sim$10 & -0.04$\pm$0.11 & +0.56$\pm$0.12 \\
IGR~J18259-0706 & Core & BPL &  - & $\sim$15 & 0.77$\pm$0.03 & 0.03$\pm$0.09 \\
QSO~B0241+62 & Core & BPL & - & $\sim$15 & -1.12$\pm$0.15 & 0.31$\pm$0.03 \\
 & Blob & - & - & $\sim$10 & 0.57$\pm$0.13 & 2.4$\pm$0.2 \\
NGC~1142 & d & - & +0.97$\pm$0.04 & - & - & -\\
 & East & - & +0.76$\pm$0.09 & - & - & - \\
 & A1 & - & +0.94$\pm$0.11 & - & - & - \\
 & A2 & - & +1.25$\pm$0.25 & - & - & - \\
 & b (Core) & BPL & - & $\sim$15 & -0.09$\pm$0.05 & +0.74$\pm$0.11 \\
\hline
NGC~4593 & Core & C- & - & $\sim$15 & -1.01$\pm$0.15 & +0.8$\pm$0.2 \\
 & Excess & - & -0.54$\pm$0.12 & - & - & - \\
IGR~J13091+1137 & Core & C- & - & $\sim$10 & -0.57$\pm$0.16 & +2.6$\pm$0.2 \\
 & Excess & - & +0.7$\pm$0.4 & - & - & - \\
IGR~J23524+5842 & Core & C- & - & $\sim$9 & -0.57$\pm$0.17 & +4.6$\pm$0.3 \\
 & North & - & 1.78$\pm$0.02 & - & - & - \\
 \hline
Mkn~50 & Core & S & -0.02$\pm$0.02 & - & - & - \\
NGC~5506 & Core & S & +0.85$\pm$0.02 & - & - & - \\
IGR~J17513-2011 & Core & S & +0.01$\pm$0.08 & - & - & - \\
IGR~J20186+4043 & Core & S & +0.90$\pm$0.03 & - & - & - \\
MGC+08-11-11 & Core & S & +0.78$\pm$0.09 & - & - & - \\
 & Jet & - & - & - & - & - \\
LEDA~170194 & Core & S & -0.10$\pm$0.08 & - & - & - \\
 & SW & - & & $\sim$10 & +0.14$\pm$0.24 & 1.9$\pm$0.4 \\
 & NE & - & & $\sim$10 & 0.7$\pm$0.5 & 2.0$\pm$0.9 \\
Mkn~3 & West & S & 1.22$\pm$0.07 & - & - & -  \\
    & East (Core) & - & 0.88$\pm$0.06 & - & - & - \\
Mkn~6 & North & - & +1.24$\pm$0.05 & - & - & - \\
     & South (Core) & S & +0.92$\pm$0.04 & - & - & - \\
NGC~1068 & d & - & +0.81$\pm$0.16 & - & - & - \\
        & a & - & +1.00$\pm$0.08 & - & - & - \\
        & b (Core) & S & +1.04$\pm$0.09 & - & - & - \\
        & SW & - & - & - & - & - \\
NGC~788 & Core & S & +0.19$\pm$0.06 & - & - & - \\
4C~+74.26 & Core & S & +0.23$\pm$0.13 & - & - & - \\
B3~0309+411B & Core & S & -0.4$\pm$0.1 & - & - & - \\
IGR~J23308+7120 & Core & S & +0.21$\pm$0.30 & - & - & - \\
NGC~4388 & NE (Core) & S & 0.67$\pm$0.09 & - & - & - \\
 & SW & - & - & $\sim$15 & +0.68$\pm$0.08 & +1.53$\pm$0.01 \\
 \hline
 \hline
\end{tabular}
\label{SED_fitting}
\end{table*}


\section{Discussion}

\subsection{Detection rates of radio emission}
\indent The detection rates of radio emission we find here are generally higher with respect to values reported in literature for surveys of Seyfert galaxies, especially at lower frequencies, typical percentages 70$-$90 \citep[e.g.][]{vanderHulstCraneKeel981,AntonucciBarvainis1988,Kellerman1989,UlvestadWilson1989,Kukula1995,RoyColbertWilsonUlvestad1998,NagarWilsonMulchaeyGallimore1999,Thean2000, Smith2020, silpa20, sebastian20, nyland20, baldi21b}. However, a comparison with these works is not straightforward for several reasons: i) the different selection criteria on which the various samples are based, which may led to selection biases; ii) the different resolution of the observations, which may result in comparison of different components because of the mismatch in the spatial scale; iii) the different frequency coverage, as at different frequencies emission from different components is expected to dominate, and the spectral shape may change; iv) the different sensitivity, as pre-upgrade observations (like the ones concerning work cited before) were characterised by higher RMS with respect to post-upgrade observations\footnote{The post-upgrade VLA can reach 4 \mujybm{}
(1-$\sigma$, in 1 hour) continuum sensitivity at most bands, one order of magnitude lower than pre-upgrade VLA \citep[e.g.][]{post_upgrade_VLA}} (like the ones presented here), which translates into different flux limits and therefore different detection rates. A proper comparison, although only at higher frequencies, can be carried out with the 22 GHz, 1-arcsec characterisation of SWIFT/BAT hard-X-ray selected AGN made by \citet{Smith2016,Smith2020}, in which they report a high ($\sim$ 96/100) detection rate, compatible with our estimate within the uncertainty at the same frequency.

As reported in Paper I, the detection rate of radio emission at high frequencies is larger than what has been reported in recent studies about mm emission in a handful of RQ AGN \citep[e.g.][]{Behar2015,Behar2018}. Moreover, if we compare with the results obtained for LLAGN only, then our detection rates are larger not only at the higher frequencies \citep[e.g.][]{Doi2011}, but also at the lower ones \citep[see for comparison e.g.][]{HU01,Saikia2018,Chiaraluce2019}.


\subsection{Morphology}

According to the morphological classification presented in the previous section, we find a prevalence of compact sources, either unresolved/slightly resolved (17/28,$\,\sim$61 per cent), with the remaining sources exhibiting jet-like features, as well as triple and more complex morphologies. 

Previous surveys on Seyfert galaxies have found morphologies in similar proportions, in which however the relative fraction depends on the frequency band, as the detection fraction of extended structures is higher at low frequencies, while the fraction of compact ones increases at higher frequencies \citep[e.g.][]{UlvestadWilsonSramek1981,vanderHulstCraneKeel981,Kukula1998,NagarWilsonMulchaeyGallimore1999,Leipski2006,Baldi2018,baldi21a}.  \citet{Gallimore2006Aug} found that, for a sample of Seyfert galaxies observed at 5 GHz with compact (D) configuration of the VLA (therefore lower angular resolution $\sim$15-20 arcsec) the fraction of sources exhibiting kiloparsec-scale radio structures (KSRs) is significantly higher ($\sim$45 per cent). They suggest that the lower fraction of extended sources found in a number of works \citep[e.g.][]{UlvestadWilson1989,Kukula1995,SchmittUlvestadAntonucciKimney2001,Thean2001} may be due to the high resolution, which resolves out the extended, low-surface brightness, steep-spectrum emission, with the KSRs being likely associated to the nuclear AGN jet activity. For our sources presented here, it is not possible to rule out that in a number of them low-surface brightness emission, may be present as well. 

In \citet{Smith2016,Smith2020}, a compact component in all the detected sources is found. In more than a half (55/96) of the sample, this is the only component (either unresolved or slightly resolved), in agreement with our findings, while in 11/96 sources jet-like structures on sub-kpc to kpc scales are present. However, they find a significant number of sources (30/96) with a diffuse morphology (i.e. extended but non-linear), which they attribute to nuclear star formation. 

\subsection{Radio nuclear spectral energy distribution}

The radio spectral slope in RQ AGN is likely the result of the combination of different emitting components dominating at different frequencies. Star-formation typically produces a synchrotron steep spectrum in the 1-10 GHz regime, while a compact core is characterized by a flat or inverted spectrum \citep{chen22}. An AGN wind/outflow and an extended jet produce an optically-thin synchrotron steep spectrum \citep{Panessa2019}.

In our sample, we find a prevalence of single power-law spectra, with the number of flat spectra equal to the number of steep spectra. If we include also single power-law sources with only two data points, then the flat spectra are 9 while the steep ones are 11. 

\begin{figure*}
\centering
\includegraphics[scale=0.35]{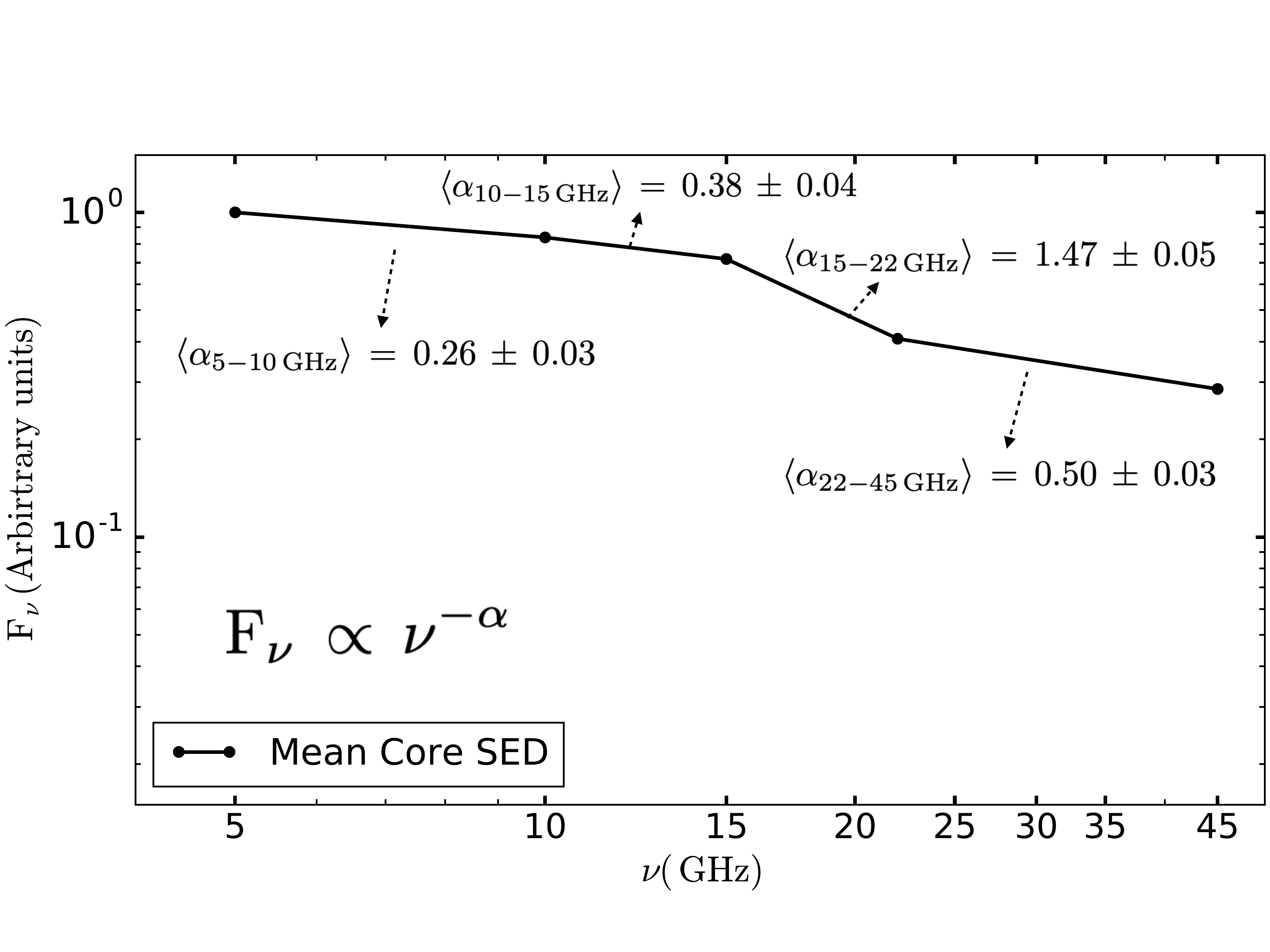}
\caption{Mean nuclear Spectral Energy Distribution in F$_{\nu}$ between 5 and 45 GHz obtained by calculating the mean of the two frequencies spectral slopes of the cores reported in Table~\ref{spectral_indices} and plotted in arbitrary units.}
\label{mean_SED}
\end{figure*}

In Fig.~\ref{mean_SED} we have plotted the mean spectral energy distribution obtained by calculating the mean of the slopes of the cores between two frequencies from Table~\ref{spectral_indices} represented in F$_{\nu}$ in arbitrary units. The average trend is that of a nearly flat slope that steepens with increasing frequency, a major steepening is observed between 15 and 22 GHz. It is interesting to note that between 22 and 45 GHz the slope flattens again. This suggests that the core emission is generally optically thick below 15 GHz, in agreement with the results presented in \citep{LaorBaldiBehar2019} for quasars with a similar range of Eddington ratios. It becomes optically thin above 15 GHz and between 22 and 45 GHz it becomes optically thick again. If we assume that the emission is produced by synchrotron, the size of the emitting region at 15 GHz is smaller than the broad line region size (from eq.22 in \citet{LaorBehar2008}). At higher frequencies, another emitting source should be even smaller, as higher frequency emission originates at smaller radii, and the flat spectrum can be the result of the the emissivity integrated over the emitting volume.
An increased radio emission is indeed expected if the observed 'mm-excess' component is extrapolated down to 45 GHz \citep{LaorBehar2008, Behar2018}.

\subsection{Radio - X-ray correlations}

It has been found that both XRB and AGN follow tight X-ray versus radio luminosity correlations, with two different tracks, i.e. one characterised by a slope of $\sim$ 0.8 and followed by low-hard state XRB and low-luminosity AGN \citep[LLAGN, e.g.][]{Gallo2003,Salvato2004,Panessa2007,Chiaraluce2019}, and one characterised by a slope of $\sim$ 1.4 and followed by 'outliers' XRB and bright RQ AGN \citep[e.g.][]{Coriat2011,Panessa2015,DongWuCao2014}. The scaling with the black-hole mass has led to the formulation of the 'fundamental' plane of black-hole activity \citep[e.g.][]{Merloni2003,Bonchi2013,DongWuCao2014}, and to theories unifying the accretion-ejection physics of XRB and AGN \citep[e.g.][]{KordingJesterFender2006}. Typically, a flat slope of $\sim$ 0.8 is associated to radiatively-inefficient accretion flows \citep[e.g.][]{Falcke2004,Coriat2011,DongWu2015}, while a steep slope of $\sim$ 1.4 is associated to radiatively-efficient accretion flows \citep[e.g.][]{Coriat2011,Gallo2012,FenderGallo2014,DongWuCao2014,QiaoLiu2015}.

\begin{figure*}
\centering
\begin{tabular}{cc}
    \includegraphics[scale=0.55]{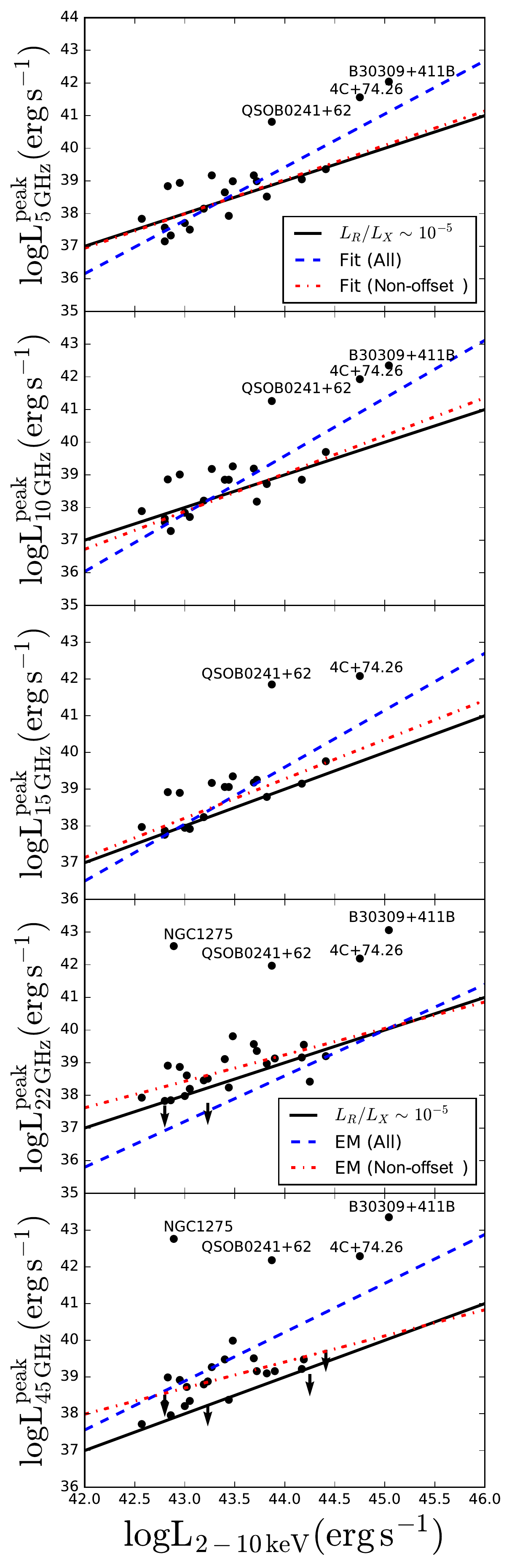} & \includegraphics[scale=0.55]{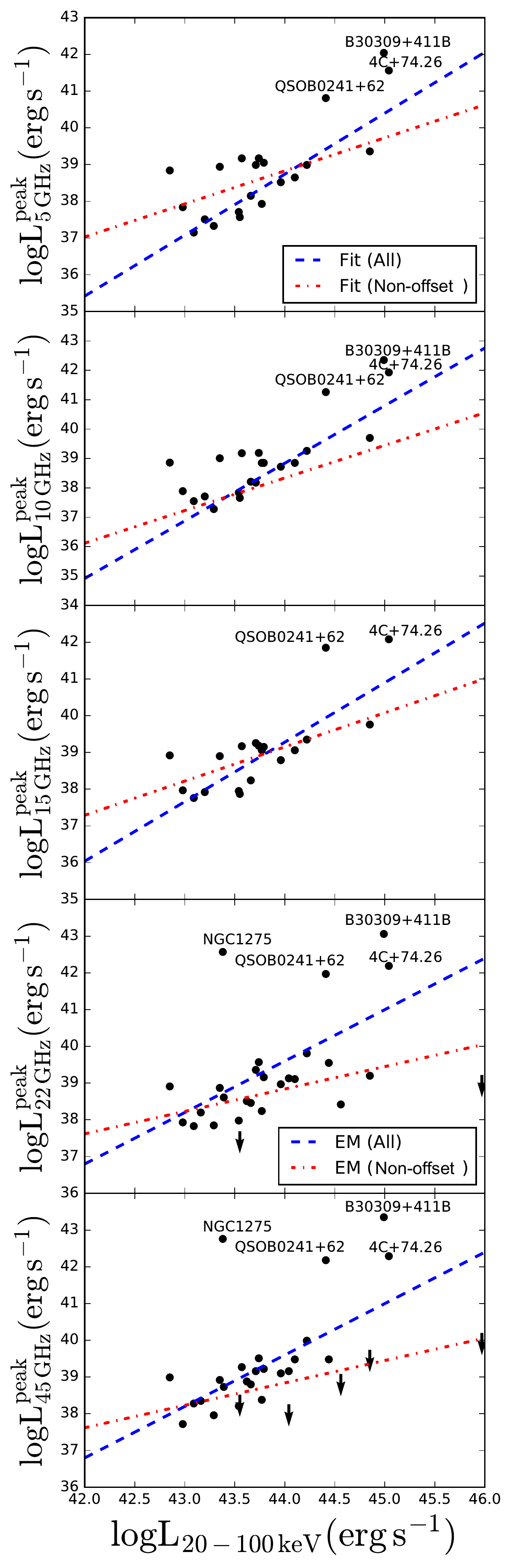} \\
\end{tabular}
\caption{Plots of the correlations between peak radio luminosities of the core components and X-ray (2-10 keV, left panel) and hard-X-ray (20-100 keV, right panel) ones. From top to bottom: 5, 10, 15, 22 and 45 GHz bands. The solid black line is the ${L_R}/{L_X}\,\sim\,10^{-5}$ relation; the dashed blue line represents a regression  considering all the data, the red dot-dashed line represents a regression discarding the offset sources (i.e. QSO~B0241+62, B3~0309+411B, NGC~1275 and 4C~+74.26, see the text). In the case of 22 and 45 GHz bands, since upper limits were present, the regression lines are calculated from a EM algorithm.}
\label{radio_xray_HXAGN_lum}
\end{figure*}

In Table \ref{relation_xray} we summarise the results of the correlations between the peak radio luminosities of the core components and the X-ray ones. In Fig. \ref{radio_xray_HXAGN_lum} we show the plots of the corresponding correlations (left panel). Since in 22 and 45 GHz bands censored data in the peak radio luminosities were present, in order to estimate the slopes we used the EM algorithm in the Astronomy Survival Analysis software package ASURV Rev 1.2 \citep[][]{Isobe1990,Lavalley1992}, which implements the methods presented in \citet{Isobe1986}. 

In order to strengthen the validity of the radio versus X-ray correlation, we have checked the significance of the corresponding flux-flux correlations. In Table 1 in Supplementary material
we report the statistical values associated to flux-flux relations, and they are plotted in Fig. 1 in Supplementary material.
The radio versus 2-10 keV flux-flux relations are strong ($\mathrm{r\,\sim\,0.6-0.8}$) and significant ($\mathrm{P\,\le\,10^{-3}}$), while in the case in which 20-100 keV fluxes are considered the significance is lower ($\sim\,10^{-2}$).

The slopes of the luminosity-luminosity relations at all frequencies (blue dashed lines in the plots), except at 10 GHz, are compatible with the slope of 1.4 found in the case of radiatively efficient sources, with high correlations coefficients and significance at 5 and 15 GHz, while at 22 and 45 GHz, where upper limits are present, the correlations are weaker and the significance is lower (order $\sim\,10^{-3}$). \\
\indent In Fig. \ref{radio_xray_HXAGN_lum}, the well known powerful radio sources previously mentioned occupy a different locus. For this reason we investigated the same correlations excluding the 'Offset' sources (red dot-dashed lines). The slopes obtained in this case tend to be flatter with respect to the previous case, with values in the range 0.6 - 1.2 and, although the correlations are strong ($r\sim$ 0.6 - 0.8), the significance of the relations is lower, i.e. $\sim10^{-3}$. This is expected, as the powerful radio sources occupy the high-luminosity region of the plane with the result of driving the overall relation \citep[e.g.][]{hardcastle09}. 

We investigated also the existence of correlations between the peak radio luminosity and hard-X-ray (i.e. 20-100 keV) one (right panel of Fig. \ref{radio_xray_HXAGN_lum}), finding strong correlations ($\rho\,=\,$0.5-0.7) with slopes in the range 1.3-2, with a high significance (order of $\sim\,10^{-3}-10^{-4}$). When the 'offset' sources are excluded, in analogy with radio-X-ray correlations, the slope flatten ($\sim\,0.6-1.1$) and, although strong ($\rho\,=\,$0.8-0.6), the significance is lower ($\sim\,10^{-2}$).


The slopes of the radio-X-ray luminosity correlations are compatible with 1.4 found by \citet{Panessa2015} using the complete sample of 79 hard X-ray selected Seyferts and 1.4 GHz data ($\sim$ 10\% of RL sources are present in their sample). However, the radio observations in \citet{Panessa2015} are at 1.4 GHz and have a resolution of $\sim$45 arcsec, much larger than ours. Therefore, the flux densities and luminosities considered for the relation may be contaminated by contribution from extended, off-nuclear component, which may be not directly AGN-related. Moreover, they consider all the 79 Seyferts in the sample of \citet{Malizia2009}, including several powerful radio sources. 
Our slopes are also consistent with the findings presented in \citet{DongWuCao2014} for a sample of bright RQ AGN with similar properties (i.e. Eddington ratios larger than 1 per cent and relatively low redshift, z $\le$ 0.3), compatible with a radiatively-efficient accretion regime as postulated in AGN-XRB unification theories. However, when powerful radio sources are excluded, the slopes become flatter. The peak of the distribution of Eddington ratios of the sources in our work is in the range between -2.25 and -1.5, while that of \citet{DongWuCao2014}, is in the range between -1 and -0.5. Our result would suggest that when intermediate accretion regimes are considered, the slope of the relation flattens, with values which are intermediate between the 0.8 slope for the very-low accretion regimes and the slope of 1.4 for higher accretion regimes.

With these observations we can test if the sources follow the empirical $\mathrm{{L_{R}}/{L_{X}}\,\sim\,10^{-5}}$ relation valid for coronally-active stars. The finding that also bright RQ AGN follow the relation has been interpreted in light of models in which both radio and X-ray would come from a hot corona \citep[e.g.][]{LaorBehar2008}. In the left panel of Fig. \ref{radio_xray_HXAGN_lum} the solid black line represents the above relation. We find that at 5 and 10 GHz the sources, except for the offset ones, roughly follow the relation, but at higher frequencies (15, 22 and 45 GHz) their peak radio luminosities tend to be systematically above the relation. In order to quantify this effect, we calculated mean values of the quantity $\log{L_{R}/L_X}$ for the non-offset sources\footnote{The offset sources, as expected, are characterised by $\log{L_{R}/L_X},\,\ge\,-3$}, considering different radio luminosities, with the \textsc{KMESTM} routine in the \textsc{ASURV} package\footnote{The routine \textsc{KMESTM} gives the Kaplan-Meier estimator for the distribution function of a randomly censored sample}. The mean values are -4.9$\pm$0.1 at 5 GHz, -4.8$\pm$0.1 at 10 GHz, -4.6$\pm$0.1 at 15 Ghz, -4.8$\pm$0.1 at 22 GHz and -4.5$\pm$0.1 at 45 GHz\footnote{In the case in which the smallest value is an upper limit the mean may suffer from statistical biases, for details see the ASURV documentation.}, with an increase of $\mathrm{\log{L_{R}/L_X}}$ ratio with frequency.
We note that the $\mathrm{\log{L_{R}/L_X}}$ relation has been found considering the 5 GHz luminosity \citep[e.g.][]{LaorBehar2008}, while \citet{Behar2018} found that when high frequencies (i.e. at $\sim$ 100 GHz) are considered, the relation rather follow a $\sim$ $10^{-4}$ trend. Indeed, we find that as the frequency increases, the sources tend to cluster in this intermediate regime. 


\begin{table*}
\centering
\caption{Results of the correlation between peak radio luminosities of the core components and the 2-10 keV luminosities, for all the sample and excluding the offset sources. Associated flux-flux correlations are also reported. \textit{Columns:} (1) the relation; (2) sample considered; (3) the slope of the relation (log-log); (4) the intercept; (5) Spearmnan's rank correlation coefficient; (6) probability of the null hypothesis that null hypothesis is that two sets of data are uncorrelated; (7) Kendall's tau; (8) probability of the null hypothesis. \\ (*) For K and Q bands, since upper limits in the dependent variables were present, the reported values are from the EM regression algorithm, the Spearman's rho and the Generalised Kendall's tau calculated in the ASURV package.}
\footnotesize
\begin{tabular}{cccccccc}
\hline
Relation & Sample & m & q & $\rho$ & P-value & $\tau$ & P-value \\ 
 & & (slope) & (intercept) & (Spearman) & (Spearman) & (Kendall's) & (Kendall's) \\
 (1) & (2) & (3) & (4) & (5) & (6) & (7) & (8) \\ 
 \hline
 \hline
 \smallskip
${\log{L_{5\,GHz}^{peak}}}-{\log{L_{2-10\,keV}}}$ & All & 1.63$\pm$0.23 & -32$\pm$10 & 0.8 & 9$\times10^{-6}$ & 0.64 & 5$\times10^{-5}$\\
 & RQ & 1.05$\pm$0.25 & -7$\pm$11 & 0.7 & 1$\times10^{-3}$ & 0.5 & 2$\times10^{-3}$\\
${\log{F_{5\,GHz}^{peak}}}-{\log{F_{2-10\,keV}}}$ & All & 1.33$\pm$0.42 & -4$\pm$5 & 0.6 & 7$\times10^{-3}$ & 0.4 & 6$\times10^{-3}$\\
 & RQ & 1.33$\pm$0.40 & -4$\pm$5 & 0.6 & 7$\times10^{-3}$ & 0.4 & 7$\times10^{-3}$\\
${\log{L_{10\,GHz}^{peak}}}-{\log{L_{2-10\,keV}}}$ & All & 1.77$\pm$0.24 & -38$\pm$10 & 0.7 & 2$\times10^{-4}$ & 0.6 & 3$\times10^{-4}$\\
 & RQ & 1.16$\pm$0.25 & -12$\pm$11 & 0.6 & 1$\times10^{-2}$ & 0.4 & 1$\times10^{-2}$\\
${\log{F_{10\,GHz}^{peak}}}-{\log{F_{2-10\,keV}}}$ & All & 1.36$\pm$0.42 & -3$\pm$5 & 0.64 & 2$\times10^{-3}$ & 0.5 & 3$\times10^{-3}$\\
 & RQ & 1.37$\pm$0.40 & -3$\pm$5 & 0.6 & 2$\times10^{-3}$ & 0.5 & 3$\times10^{-3}$ \\
${\log{L_{15\,GHz}^{peak}}}-{\log{L_{2-10\,keV}}}$ & All & 1.55$\pm$0.30& -29$\pm$11 & 0.8 & 2$\times10^{-5}$ & 0.6 & 1$\times10^{-4}$\\
 & RQ & 1.07$\pm$0.20 & -8$\pm$9 & 0.76 & 3$\times10^{-4}$ & 0.6 & 1$\times10^{-3}$\\
${\log{F_{15\,GHz}^{peak}}}-{\log{F_{2-10\,keV}}}$ & All & 1.16$\pm$0.40 & -5$\pm$5 & 0.6 & 7$\times10^{-3}$ & 0.4 & 9$\times10^{-3}$\\
 & RQ & 1.15$\pm$0.44 & -5.4$\pm$4.9 & 0.6 & 7$\times10^{-3}$ & 0.43 & 9$\times10^{-3}$ \\
  \hline
${\log{L_{22\,GHz}^{peak}}}-{\log{L_{2-10\,keV}}}^{*}$ & All & 1.4$\pm$0.4 & -23$\pm$17 & 0.64 & 1$\times10^{-3}$ & z$\sim$3.5 & 4$\times10^{-4}$ \\
 & RQ & 0.81$\pm$0.21 & 3.6$\pm$9.0 & 0.7 & 1.4$\times10^{-3}$ & z$\sim$3.2 & 1.5$\times10^{-3}$ \\
${\log{F_{22\,GHz}^{peak}}}-{\log{F_{2-10\,keV}}}^{*}$ & All & 1.4$\pm$0.5 & -3$\pm$5 & 0.6 & 2$\times10^{-3}$ & z$\sim$3.2 & 1$\times10^{-3}$ \\
 & RQ & 1.05$\pm$0.30 & -6.8$\pm$3.0 & 0.6 & 5$\times10^{-3}$ & z$\sim$3.17 & 1.5$\times10^{-3}$ \\
${\log{L_{45\,GHz}^{peak}}}-{\log{L_{2-10\,keV}}}^{*}$ & All & 1.33$\pm$0.40 & -18$\pm$17 & 0.6 & 3$\times10^{-3}$ & z$\sim$3.4 & 7$\times10^{-4}$ \\
 & RQ & 0.71$\pm$0.2 & 8$\pm$8 & 0.6 & 6$\times10^{-3}$ & z$\sim$3.04 & 2$\times10^{-3}$\\
${\log{F_{45\,GHz}^{peak}}}-{\log{F_{2-10\,keV}}}^{*}$ & All & 1.4$\pm$0.5 & -3$\pm$6 & 0.6 & 2$\times10^{-3}$ & z$\sim$3.3 & 1$\times10^{-3}$ \\
 & RQ & 1.0$\pm$0.3 & -7.4$\pm$3.0 & 0.6 & 4$\times10^{-3}$ & z$\sim$3.23 & 1$\times10^{-3}$ \\
\hline
\hline
\end{tabular}
\label{relation_xray}
\end{table*}


There is still no definite explanation for the observed functional form of relation and care should be taken when deriving physical conclusions based on them. Indeed, several works reported that different slopes are found when considering different classes of objects, i.e. only RL sources \citep[][]{Wang2006}, Giga-Hertz Peaked and Compact Steep Spectrum sources \citep[GPS and CSS, e.g.][]{FanBai2016}, Narrow-Line Seyfert 1 \citep[NLS1, e.g.][]{YaoQiaoWuYan2018}, and blazars \citep{ZhangZhangZhang2018}.
Magnetically dominated coronae may explain the observed relations in RQ AGN \citep[e.g.][and references therein]{LaorBehar2008}. Alternatively, an interpretation in terms of a coupling of inflow and outflow as in XRBs may apply \citep[e.g.][and references therein]{carotenuto21}.

\section{Clues on the origin of nuclear radio emission}

Four sources in our sample are well known powerful radio emitters QSO~B0241+62, B3~0309+411B, NGC~1275 and 4C~+74.26. 
Their radio properties are consistent with the jet activity already reported by previous studies \citep[e.g.][]{Hutchings1982,deBruyn1989,Pedlar1990,HU01,Healey2007,Lister2019,Bruni2020}.

The remaining sources (26/30) can be classified as RQ considering their radio properties. Two of them are undetected (IGR~J16426+6536 and IGR~J18027-1455) and one of them (2E~1853.7+1534) has been detected at 22 GHz (only 22 and 45 GHz observation were available) resulting in a diffuse component that maybe consistent with jet/outflow or starburst emission.

Steep spectra are found in 11/30 sources. Five of them, namely NGC~1068, MCG+08-11-11, Mkn~3, Mkn~6 and NGC~4388, exhibit resolved, jet-like structure, sometimes coupled to more complex features. For these sources, the origin of observed radio emission is compatible with optically-thin synchrotron emission from a sub-relativistic jet. This in agreement with previous studies reporting aligned structures on mas scales, high-brightness temperature ($\mathrm{\gg\,10^8\,K}$) and flat spectrum for the mas cores \citep[e.g.][]{Kukula1996,Kukula1999,Lal2004}. However, in NGC~1068 and NGC~4388, while a jet is observed at kpc and sub-kpc scales, mas scales observations revealed low brightness temperature cores ($\mathrm{\sim\,10^5\,K}$), compatible with thermal free-free emission from the inner torus \citep[e.g.][]{Gallimore1996Feb,Kukula1999,Mundell2000,GirolettiPanessa2009}. Six sources, namely NGC~4151, IGR~J00333+6122, LEDA~168563, NGC~5506, IGR~J1638-2057 and IGR~J20186+4043, exhibit a compact morphology and a steep spectrum. In this case, the observed radio emission may be due to optically-thin emission from a sub-relativistic jet which is unresolved by our observations, as already observed in NGC~4151 and NGC~5506 \citep[e.g.][]{Pedlar1993,Mundell2003,Middelberg2004,Ulvestad2005}. Radio emission from star formation may instead occur at larger (a few kpc) scales \citep[e.g.][]{Padovani2011,Condon2013}. For the remaining sources optically-thin synchrotron emission from disc winds can not be ruled out and higher resolution observations are needed. \\
\indent Three sources, namely IGR~J23524+5842, NGC~4593 and IGR~J13091+1137, exhibit peaked spectra in the GHz range. The source IGR~J23524+5842 exhibits a core coupled to a one-sided lobe, and its radio emission likely comes from a jet; for the other two this hypothesis can not be confirmed, as they are compact and the SED exhibits an additional component at high frequencies. \\
\indent Two more sources, namely NGC~5252 and NGC~1142, exhibit cores with flat spectra up to a break (at $\sim$10 and $\sim$15 GHz, respectively). This emission is likely due to a jet, the subsequent spectral decline is probably marking the transition to the optically-thin regime. In the former, the non-thermal origin of radio emission has been confirmed by VLBI studies \citep[e.g.][]{Mundell2000}, while in the latter this scenario cannot be confirmed. The extra-nuclear radio emission of NGC~1142 is of star formation origin and it is discussed in a dedicated section in Appendix A. \\
\indent Finally, seven sources, namely NGC~788, 4U~0517+17, Mkn~50, LEDA~170194, IGR~J17513-2011, IGR~J18259-0706 and IGR~J23308+7120, exhibit compact flat-spectrum cores. In addition to the flat core,  LEDA~170194 shows a steep-spectrum two-sided jet extending over $\sim$2 kpc. \\
\indent The observed radio sources above exhibit flat spectra up to 45 GHz (with the exception of IGR~J23308+7120, which is flat up to 15 GHz and undetected at 22 and 45 GHz). This may be due to optically-thick synchrotron emission from a compact jet. However, flat radio spectra extending up to high frequencies, may also be due to synchrotron radio emission from a magnetically-heated corona \citep[e.g.][]{LaorBehar2008}. In the so-called coronal models, the corona is a radially-stratified plasma, both in the magnetic field and number density of relativistic electrons, in either a spherical or disc geometry. Each plasma shell emits as optically-thick synchrotron radiation, but variations in the opacity from one shell to another produce different turnover frequencies, such that external shells dominate low frequencies, while inner ones dominate the high frequency range. The result is an overall flat radio spectrum \citep[e.g.][]{RaginskiLaor2016}. One difference between the compact jet and coronal emission scenario is that in the former a break in the spectral slope (due to the optically-thick/optically-thin transition) is expected at relatively low frequencies (a few tens of GHz), while the compactness of the corona can result in flat spectra up to $\sim$ 300 GHz \citep[e.g.][]{LaorBehar2008,RaginskiLaor2016}. This suggests that, with the resolution and frequency coverage of our observations, the external layers of the corona are eventually sampled. 

In order to put more stringent constraints to the origin of radio emission in these sources VLBI observations will be fundamental. Indeed, it will be possible to determine the mas size of the sources, as a size smaller than $\sim$pc rules out star-formation or an AGN wind interacting with the interstellar medium. Moreover, it will also exclude free-free emission, as a brightness temperature much larger than $\mathrm{\sim\,10^{6}\,K}$ is compatible only with non-thermal emission \citep[e.g.][]{CondonRansom2016}. Finally, variability studies can constrain the size of the emitting regions and therefore discriminate between star formation and non-thermal emission, like the base of a jet or a corona, expected to vary on relatively small timescales \citep[e.g.][]{Barvainis2005,Baldi2015, panessa22}.


\section{Conclusions}

In this paper we presented wide-band (5 - 45 GHz), high-sensitivity (a few $\mathrm{{\mu}Jy\,beam^{-1}}$), sub-arcsec (1-0.4\arcs) JVLA observations for a sample of 30 nearby ($0.003\,\le\,z\,\le\,0.3$) and moderately accreting ($\mathrm{-2.5\,\le\,\log{L/L_{Edd}}\,\le\,-0.5}$) hard-X-ray selected AGN. The observations allow to characterise the sub-kpc scale radio properties of the sample and build wide-band SEDs, with the final aim of investigating the origin of radio emission. Below we summarise our main findings:

\newenvironment{myitemize}
{ \begin{itemize}
    \setlength{\itemsep}{0pt}
    \setlength{\parskip}{0pt}
    \setlength{\parsep}{0pt}     }
{ \end{itemize}                  }

\begin{myitemize}
    \item We find a high detection fraction of radio emission at all frequencies, that decreases with frequency (21/21, 21/21, 19/19, 26/29 and 24/29, i.e. 96, 96, 95, 87 and 81 per cent at 5, 10, 15, 22 and 45 GHz, respectively). These values occupy the high tail of the distribution of the detection rates reported in literature for RQ AGN, typically around 70 - 90 per cent.
    \item The radio luminosities of the detected sources are in the range between 37 $\,\le\,\log{L_{R}(erg\,s^{-1}})\,\le$ 40. 
    Four sources are known powerful radio emitters and indeed show larger luminosities $\ge$ 41 (in log units).
    \item Two sources (observed only at 22 and 45 GHz) have not been detected. The upper limits on their flux densities are $\sim$0.15 and 0.1 $\mathrm{mJy\,beam^{-1}}$. One of them, IGR~J16426+6536, has been observed twice, $\sim$17 months apart, in both cases it is undetected. The upper limits on radio luminosities imply values of the X-ray radio-loudness $R_X\,<-$4.5, at the very-low-power tail of radio-loudness distribution, possibly hinting to a 'radio-silent' nature.
    \item The morphology of the sources varies with the selected frequency, where extended emission dominates the lower frequencies. Overall, the majority of them are compact, characterised by a core component (17/30). The remaining sources exhibit either one-sided (7/30) or two sided (1/30) jets. Three sources exhibit a complex morphology.
    \item The SED of the sources can be classified into three categories: single power-law spectra (14/30), convex spectra (3/30) and broken power law spectra (4/30). For 6/30 sources only 2 SED data points (at 22 and 45 GHz) are available, while for one source only one data point is available. We derive a mean radio spectral energy distribution of the cores.
    \item We find a significant radio-X-ray correlation at all frequencies. When the four powerful radio emitters are excluded, the slopes of the correlation range between $\mathrm{m=0.7-1.2}$. This range is intermediate between the slope of 1.4 found for high luminosity AGN and the slope of 0.8 found for LLAGN, in agreement with the intermediate accretion rates and luminosities covered by the sample considered here.
    \item Our sources roughly follow the empirical Gudel-Benz relation, with a transition from the $\mathrm{10^{-5}}$ to the $\mathrm{10^{-4}}$ relation at higher frequencies (i.e. 15, 22 and 45 GHz), suggesting a possible contribution to both radio and X-ray from a hot corona. The four powerful radio emitters significantly depart from the relation, as expected.
 
 \end{myitemize}

The combination of morphology, SED shape and (when available) VLBI information from the literature allows to formulate possible scenarios for the radiative mechanisms responsible for the radio emission: i) $\sim$13\% of the sample is RL; ii) $\sim$37\% of the sample has steep spectra compatible with optically-thin synchrotron from a jet, which is unresolved in compact cores. In $\sim$13\% of them, optically-thin synchrotron from disc winds can not be ruled out; iii) flat spectrum sources are generally compact (except for LEDA~170194 which exhibits a two-sided jet), indicating a possible optically-thick synchrotron emission from a compact jet and/or a hot corona ($\sim$30\%); iv) in $\sim$30\% of the sample, peaked core spectra are found (in one case showing a one-sided jet morphology) again suggesting a possible jet scenario.
    
This work is part of a large project aiming at investigating the origin of radio emission through high-resolution, high-sensitivity, multi epoch, wide-band radio observations. Our volume limited, well characterized hard X-ray sample is perfectly suited to push over this effort by exploiting the unprecedented capabilities of the Square Kilometer Arrays (SKA) pathfinders and precursors.


\section*{Acknowledgements}
We acknowledge the anonymous referee for her/his valuable suggestions and comments that have improved the quality of our manuscript. The authors would like to thank the NRAO helpdesk staff for the help in the data reduction. The National Radio Astronomy Observatory is a facility of the National Science Foundation operated under cooperative
agreement by Associated Universities, Inc. EC acknowledges the National Institute of Astrophysics (INAF) and the University of Rome – Tor Vergata for the PhD scholarship in the XXXIII PhD cycle for the period December 2017 - December 2020.
GB and FP acknowledge financial support under the INTEGRAL
ASI-INAF agreement 2019-35-HH.0. Part of the data reduction has been performed by EC at the DiFA - University of Bologna and the Istituto di Radioastronomia in the period January - March 2020, a visit funded by the Bando Fondi Ricerca 2019 of the INAF-IAPS. FP acknowledges support from a
grant PRIN-INAF SKA-CTA 2016. AL was supported by the Israel Science Foundation (grant no. 1008/18). E.B. is supported by a Center of Excellence of the Israeli Science Foundation (grant No. 2752/19). IMcH thanks STFC for support under grant  ST/R000638/1. FT acknowledges support by the Programma
per Giovani Ricercatori - anno 2014 Rita Levi Montalcini.
This research made use of APLpy, an open-source plotting
package for Python \citep{aplpy}.

\section*{Data Availability} 

The data underlying this article were accessed from NRAO (https://science.nrao.edu/). The derived data generated in this research will be shared on reasonable request to the corresponding author.


\bibliographystyle{aa}
\bibliography{ref} 

%
%
%

\appendix

\section{NGC~1142}

NGC~1142 is a massive spiral galaxy and is part of an interacting system \textit{Arp~118}, in which the companion is the massive elliptical galaxy NGC~1143, $\sim$40 arcsec far away in the NW direction ($\sim$24 kpc). It is most probably the result of a collisional encounter \citep[e.g.][]{JoyGhigo1988}.

Its radio map exhibits a complex morphology, with a central component, named B, where the core is believed to reside, plus additional extra-nuclear components on scales of $\sim$10 kpc (total extension). At 5, 10 and 15 GHz, a southern ridge of radio emission extending on approximately the same spatial scale is visible. At 22 GHz 6 components are detected plus a southern ridge, namely East, A1, A2, C, B and D. In the 45 GHz naturally-weighted map only three components are detected, namely A1, A2 and B. This is due to a combined effect of the steep spectrum of the off-nuclear components and the lower sensitivity at 45 GHz with respect to 22 GHz. At 5 GHz (archival post-upgrade JVLA observations) the East component is blended with A1, while the C component is not detected. At 10 and 15 GHz the significance of the detection of the off-nuclear components is lower. This is due to a combined effect of the steep spectrum and the higher RMS in these bands with respect to 5 and 22 GHz maps (archival VLA observations were of lower sensitivity before the upgrade). The observed morphology is compatible with previous radio studies at comparable resolution \citep[e.g.][]{JoyGhigo1988,Condon1990,Thean2000}, although an additional component, C, is detected with a high significance at 22 GHz.

The present observations at matched resolution, although with slightly different sensitivities, allow to build for the first time the SED for all the detected components. The core (component B) exhibit a spectrum which is flat/inverted up to to a break occurring at $\sim$15 GHz, with a steep decline after it (see Table \ref{SED_fitting}). This likely originates from the base of a jet, However, due to the missing information at the mas scale, it is not possible to rule out a possible thermal origin.

The extra-nuclear components are characterised by steep spectra in cm/sub-cm range, with spectral indices in the range 0.7-1.3 (see Table \ref{SED_fitting} and bottom-right panel of Fig. \ref{fig:4classes}). These values, together with the observed morphology, are compatible with radio emission from star formation \citep[e.g.][]{Condon1992,Panessa2019}. This source is part of the Calar Alto Legacy Integral Field spectroscopy Area (CALIFA) survey \citep[][]{CALIFA} which covers
the range of 3700 - 7000 angstrom with a 3.7 arcsec resolution.

\begin{figure*}
\scriptsize
\centering
\includegraphics[scale=0.775]{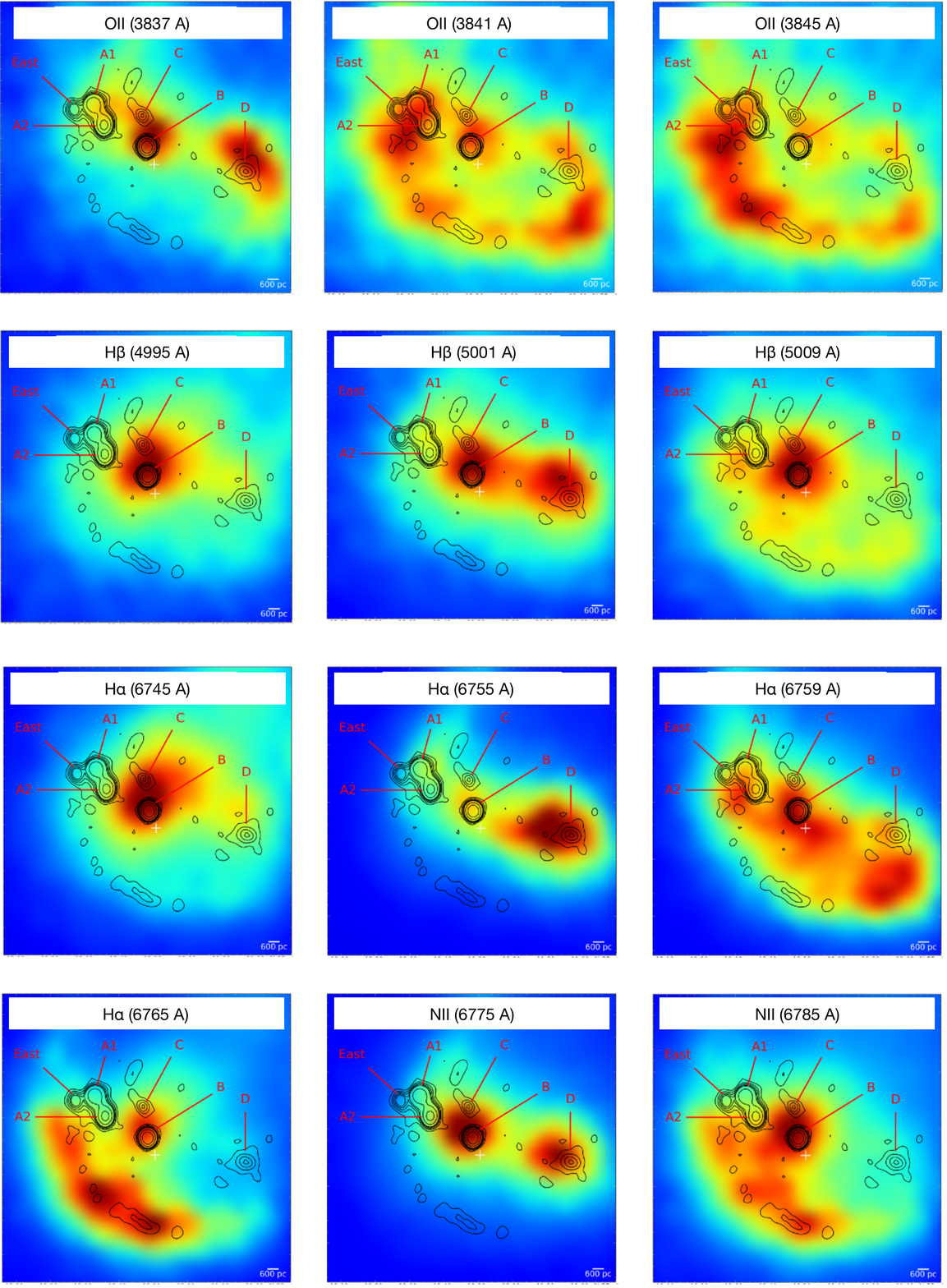}
\caption{Maps for NGC~1142 in which we overplot our proprietary K-band contours to CALIFA emission lines maps at different wavelengths, comprising OII, H$\beta$, H$\alpha$ and NII. In each map, the wavelength corresponding to the optical emission line image is indicated (in angstroms).}
\label{fig:NGC1142_CALIFA}
\end{figure*}

In Fig. \ref{fig:NGC1142_CALIFA} we show over-plots of 22 GHz contours with optical emission lines images corresponding to different lines, i.e. NII, OII, H{$\alpha$} and H${\beta}$ (the corresponding wavelengths of each optical map are indicated on top). These lines may trace recent star formation \citep[e.g.][]{Calzetti2007}.
From Fig. \ref{fig:NGC1142_CALIFA} it is possible to make the following considerations: i) the component B, the nucleus, is always optically bright, as expected; ii) the component D is bright in all emission lines; iii) the southern ridge is bright in NII, OII and H$\alpha$, it is weak in H$\beta$; iv) the triple group East/A1/A2 is bright in OII, it is weak in H$\alpha,\beta$ and NII. 

The above considerations are suggesting that the radio emission from the extra-nuclear steep spectrum components, coinciding with spikes in optical emission lines maps, is indeed synchrotron radiation from regions of intense star formation, as first proposed by \citet{JoyGhigo1988}. Previous works support this interpretation and have found that the component D indeed coincides with a giant HII region \citep[e.g.][]{Hippelein1989}; the southern ridge is part of the ring of the disk galaxy which is stripped in an elliptical geometry because of encounter with the companion galaxy, which has produced an enhancement in the molecular gas density, as traced by CO(1-0) observations \citep[e.g.]{Gao1997}. In this scenario, the coincidence of star forming regions traced by optical and radio observations with regions of increased gas density is supported by dynamical models \citep[e.g.][]{Lamb1998}. While the interpretation of the origin of steep-spectrum radio emission of the extra-nuclear components D and southern ridge fits with the above picture, that of triple East/A1/A2 is not straightforward, as it appears weak in emission lines (except for OII) and it is spatially coincident with a gap in the molecular gas CO(1-0) \citep{Gao1997}. Interpretations for the origin of this radio emission comprise emission from SNR in molecular gas clouds with low density \citep[e.g.][]{Gao1997}, radio emission from star formation triggered by the temporary enhanced density in the impact locus \citep[e.g.][]{Lamb1998} or emission from cosmic rays from star formation and supernovae explosions \citep[e.g.][]{Appleton2003}.


\end{document}